\documentclass[fleqn,usenatbib]{mnras}

\usepackage[T1]{fontenc}

\DeclareRobustCommand{\VAN}[3]{#2}
\let\VANthebibliography\thebibliography
\def\thebibliography{\DeclareRobustCommand{\VAN}[3]{##3}\VANthebibliography}

\usepackage{graphicx}	
\usepackage{amsmath}	
\usepackage{amssymb}	
\usepackage{physics}
\usepackage{bm}

\renewcommand{\vec}{\bm}

\title[Stellar Clustering and Orbital Misalignments]{The Impact of Stellar Clustering on the Observed Multiplicity of Super-Earth systems: Outside-in Cascade of Orbital Misalignments Initiated by Stellar Flybys}

\author[Laetitia Rodet \& Dong Lai]{
Laetitia Rodet,$^{1}$\thanks{E-mail: lbr63@cornell.edu}
Dong Lai,$^{1}$
\\
$^{1}$Cornell Center for Astrophysics and Planetary Science, Department of Astronomy, Cornell University, Ithaca, NY 14853, USA
}

\date{Accepted 2021 October 18. Revised: 2021 October 15; Received 2021 July 12; in original form ZZZ}

\pubyear{2021}

\begin{document}
\label{firstpage}
\pagerange{\pageref{firstpage}--\pageref{lastpage}}
\maketitle

\begin{abstract}
	A recent study suggests that the observed multiplicity of super-Earth
	(SE) systems is correlated with stellar overdensities: field stars in high phase-space density environments have an excess of single-planet systems compared to stars in low density fields. This correlation is puzzling as stellar clustering is expected to influence mostly the outer part of planetary systems. Here we examine the possibility that stellar flybys indirectly excite the mutual inclinations of initially coplanar SEs, breaking their co-transiting geometry. We propose that flybys excite the inclinations of exterior substellar companions, which then propagate the perturbation to the inner SEs. Using analytical calculations of the secular coupling between SEs and companions, together with numerical simulations of stellar encounters,
	we estimate the expected number of ``effective'' flybys per planetary
	system that lead to the destruction of the SE co-transiting geometry.
	Our analytical results can be rescaled easily for various SE and
	companion properties (masses and semi-major axes) and stellar cluster
	parameters (density, velocity dispersion and lifetime).  We show that
	for a given SE system, there exists an optimal companion architecture
	that leads to the maximum number of effective flybys; this results from
	the trade-off between the flyby cross section and the companion’s
	impact on the inner system.  Subject to uncertainties in the cluster
	parameters, we conclude that this mechanism is inefficient if the SE
	system has a single exterior companion, but may play an important role
	in ``SE + two companions'' systems that were born in dense stellar
	clusters. Whether this effect causes the observed correlation between planet multiplicity and stellar overdensities remains to be confirmed.
\end{abstract}




\section{Introduction}

Planets with masses/radii between Earth and Neptune, commonly called super-Earths (SEs) or mini-Neptunes, can be found around 30\% of solar-type stars \citep[e.g.][]{lissauerArchitectureDynamicsKepler2011,fabryckyArchitectureKeplerMultitransiting2014,winnOccurrenceArchitectureExoplanetary2015,zhu30SunlikeStars2018}. A large sample of them were discovered by the Kepler mission, with orbital periods below 300 days. SE systems contain an average of three planets \citep{zhu30SunlikeStars2018}, generally ``dynamically cold'', with eccentricities $e \sim 0.02$ and mutual inclinations $\Delta I \sim 2^\circ$ \citep[e.g.,][]{winnOccurrenceArchitectureExoplanetary2015}. However, there is an observed excess of single transiting planets, that could be sign of a dynamically hot sub-population of misaligned planets \citep[so-called Kepler dichotomy,][]{lissauerArchitectureDynamicsKepler2011,johansenCanPlanetaryInstability2012,readTransitProbabilitiesSecularly2017}, and indicating that the mutual inclinations in a multi-planet system decrease with the number of planets \citep{zhu30SunlikeStars2018,heArchitecturesExoplanetarySystems2019,millhollandEvidenceNonDichotomousSolution2021}

Recent work by \cite{longmoreImpactStellarClustering2021} has revealed an
intriguing correlation between stellar phase-space density and the architecture of planetary systems, in particular the multiplicity. This work followed a similar analysis by \cite{winter_stellar_2020}, which uncovered a correlation between stellar phase-space density and the occurrence of hot Jupiters. Using Gaia DR2 data \citep{gaia_collaboration_gaia_2018}, \cite{longmoreImpactStellarClustering2021} computed the local stellar phase-space density of planet-hosting stars and their neighbours (within 40 pc) to determine whether the exoplanet host was in a relatively low or high phase-space density zone compared to its neighbours. They hypothesized that stars in current stellar overdensities were previously part of dense stellar clusters, from which only local residual overdensities remain. They showed that Kepler systems in local stellar phase-space overdensities have a significantly larger single-to-multiple ratio compared to those in the low phase-space density environment. The origin of this correlation is puzzling, as stellar clustering is expected to affect mostly the outer part of planetary systems in very dense environments \citep{laughlin_modification_1998,malmberg_effects_2011,parker_effects_2012,cai_stability_2017,li_flyby_2020}. Recent works have also suggested that the correlation is weaker when using a smaller unbiased stellar sample   \citep{adibekyanStellarClusteringOrbital2021}, and that the current stellar overdensities could be associated to stellar age or to galactic-scale ripples as opposed to dense birth clusters \citep{mustillHotJupitersCold2021,kruijssenNotBirthCluster2021}. Alternatively, new studies have suggested that stellar flybys can excite the eccentricities and inclinations of outer planets/companions, which then trigger the formation of hot Jupiters from cold Jupiters via high-eccentricity migration \citep{wang_hot_2020,rodetCorrelationHotJupiters2021}. For this flyby scenario to be effective, certain requirements \citep[derived analytically in][]{rodetCorrelationHotJupiters2021} on the companion property (mass and semi-major axis) and the cluster property (such as stellar density and age) must be satisfied. In this paper, we will examine a similar ``outside-in'' effect of stellar flybys on the SE systems. Earlier, \cite{zakamskaExcitationPropagationEccentricity2004} examined the excitation and inward propagation of eccentricity disturbances in planetary systems. Our work focuses on inclination disturbances, as they are most relevant in determining the co-transit geometry of multi-planet systems.
 
In recent years, long-period giant planets have been observed in an increasing number of SE systems. Statistical analysis combining radial velocity and transit observations suggests that, depending on the metallicities of their host stars, 30–60 \% of inner SE systems have cold Jupiter companions \citep{zhuSuperEarthColdJupiter2018,bryanExcessJupiterAnalogs2019}. The dynamical perturbations from external companions can excite the eccentricities and mutual inclinations of SEs, thereby influencing the observability (co-transiting geometry) and stability of the inner system \citep{boueCompactPlanetarySystems2014,carreraSurvivalHabitablePlanets2016,laiHidingPlanetsBig2017,huangDynamicallyHotSuperEarths2017,mustillEffectsExternalPlanets2017,hansenPerturbationCompactPlanetary2017,beckerEffectsUnseenAdditional2017,readTransitProbabilitiesSecularly2017,puEccentricitiesInclinationsMultiPlanet2018,denhamHiddenPlanetaryFriends2019a,puStrongScatteringsCold2021,rodetInclinationDynamicsResonant2021}. This requires the giant planets to be dynamically hot, which can be achieved by planet-planet scattering. Alternatively, in dense stellar environments, the eccentricities and inclinations of giant planets can be increased by stellar flybys.

In this paper, we examine the possibility that SEs systems become misaligned following a stellar encounter (``flyby-induced misalignment cascade'' scenario), using a combination of analytical calculations (for the secular planet interactions) and numerical simulations (for stellar flybys). In Section~\ref{sec:scenario}, we outline the proposed scenario and its key ingredients. In Section~\ref{sec:coupling}, we examine the inclination requirement for an outer companion to break the co-transiting geometry of two inner initially coplanar SEs. In Section~\ref{sec:flyby}, we evaluate the extent to which a stellar encounter can raise the inclination of an outer companion. In Section~\ref{sec:rate}, we derive the expected number of ``effective'' flybys (i.e. those that succeed in breaking the co-transiting geometry of SEs) per system as a function of the properties of the planetary system (SEs+companion) and its stellar cluster environment. Finally, in Section~\ref{sec:rate2p}, we extend our calculation to SE systems with two outer companions. In Section~\ref{sec:mass}, we examine the effect of different flyby masses. In Section~\ref{sec:conclusion}, we summarize our main results, recalling the key figures and equations of the paper.

\section{Scenario: Flyby-Induced Misalignment Cascade}
\label{sec:scenario}

The observed excess of single-transiting SEs or mini-Neptunes could be sign of a dynamically hot sub-population of inner ($\lesssim 0.5$ au) planets with appreciable mutual inclinations. The recently-evidenced correlation between single-transiting systems and stellar overdensities suggests that stellar environments could play a significant role in this misalignment. However, close-in SEs are largely protected from perturbations induced by the stellar environment thanks to their proximity to the host star. 

According to recent statistical estimates, 30--60\% of inner SE systems have cold Jupiter companions \citep[$\gtrsim 1$ au, mass $\gtrsim 0.3$ M$_{\rm J}$; see][]{zhuSuperEarthColdJupiter2018, bryanObliquityConstraintsExtrasolar2020}. Such companions widen the effective cross section of the planetary system, and thus its vulnerability to the dynamical perturbations from the stellar environment. The architecture of the inner SEs can be indirectly impacted by stellar flybys through the eccentricity/inclination excitations of the outer companions.

\begin{figure}
	\centering
	\includegraphics[width=\linewidth]{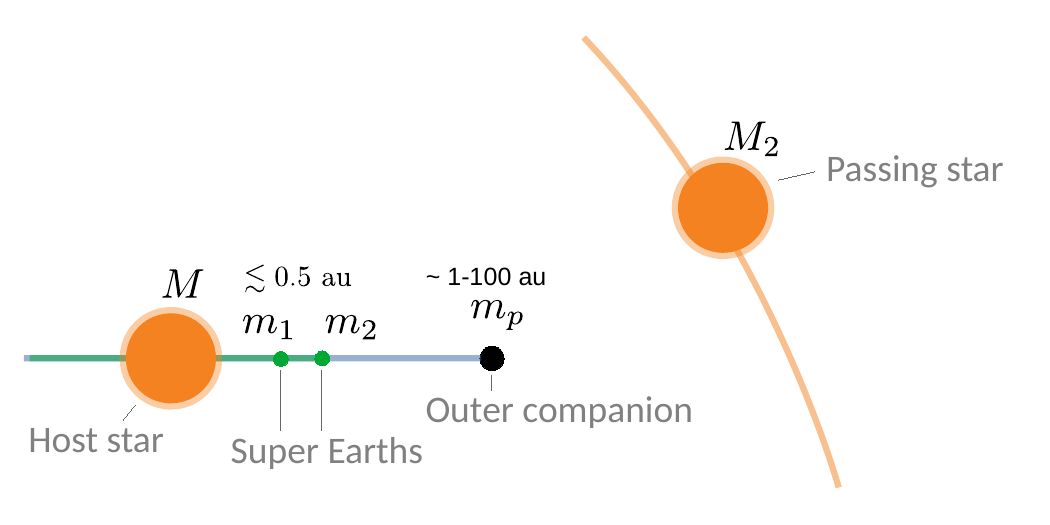}
	\caption{Schematic of the flyby-induced misalignment cascade scenario: An initially coplanar system comprising two inner planets (SEs) and an outer companion encounters a passing star. The stellar flyby changes the orbital inclination of $m_{\rm p}$, which then induces misalignment between the orbits of $m_1$ and $m_2$.}\label{fig:notations}
\end{figure}

In this paper, we consider a system of two SEs with masses $m_1$, $m_2$ and semimajor axes $a_1 <  a_2 \lesssim 0.5$ au, on circular coplanar orbits around a host star (mass $M = M_1 \sim 1~\mathrm{M}_{\sun}$, radius $R_1 \sim 1~\mathrm{R}_{\sun}$). An outer companion (mass $m_{\rm p} \ll M$) is located at $a_{\rm p}$ on an initially circular coplanar orbit (see Figure~\ref{fig:notations}). In Section~\ref{sec:rate2p} we will consider the case of two exterior companions surrounding SEs.

The system may experience close encounters with other stars, if the stellar density is large enough. This is typically the case in the beginning of the planetary system's life, when it is embedded in its dense birth cluster. Most of these clusters are loosely bound or unbound, so that they tend to be disrupted over tens of millions of years (the current phase-space overdensities would be the remnants of these clusters). However, in the early phase of the cluster, close encounters may be frequent enough to influence the planetary architecture. 

We denote by $e_{\rm p}$ and $I_{\rm p}$ the eccentricity and inclination of the companion $m_{\rm p}$ after an encounter with a passing star $M_2$. The inclined companion then induces mutual inclination within the SEs through secular interactions. In order for the SEs to break their co-transiting geometry, their relative inclination $\Delta I$ must be greater than a critical value
\begin{equation}
	\Delta I_{\rm crit} \simeq \frac{R_1}{a_2} \simeq 0.5^\circ \left(\frac{a_2}{0.5~\mathrm{au}}\right)^{-1} \left(\frac{R_1}{1~\mathrm{R_{\sun}}}\right). \label{eq:Ditransit}
\end{equation}

In the following sections, we will derive the minimum required inclination of the companion $I_{\rm p, crit}$ to break the SEs co-transiting geometry, and the corresponding likelihood that a stellar encounter would generate such an inclination.

\section{Secular Dynamics of Planet Misalignment}
\label{sec:coupling}

When the outer companion planet to the SEs in an initially coplanar system suddenly becomes misaligned by $I_{\rm p}$, the mutual inclination $\Delta I$ between the SEs will oscillate around a forced value \citep{laiHidingPlanetsBig2017,rodetInclinationDynamicsResonant2021}. In the following, we derive the minimum misalignment $I_{\rm p, crit}$ required for the forced inclination to be greater than the critical value $\Delta I_{\rm crit}$ (see equation~\ref{eq:Ditransit}). In theory however, due to the oscillation of $\Delta I$, the co-transiting geometry of the SEs will be broken only part of the time even if the forced inclination is greater than $\Delta I_{\rm crit}$.

\begin{figure}
	\centering
	\includegraphics[width=\linewidth]{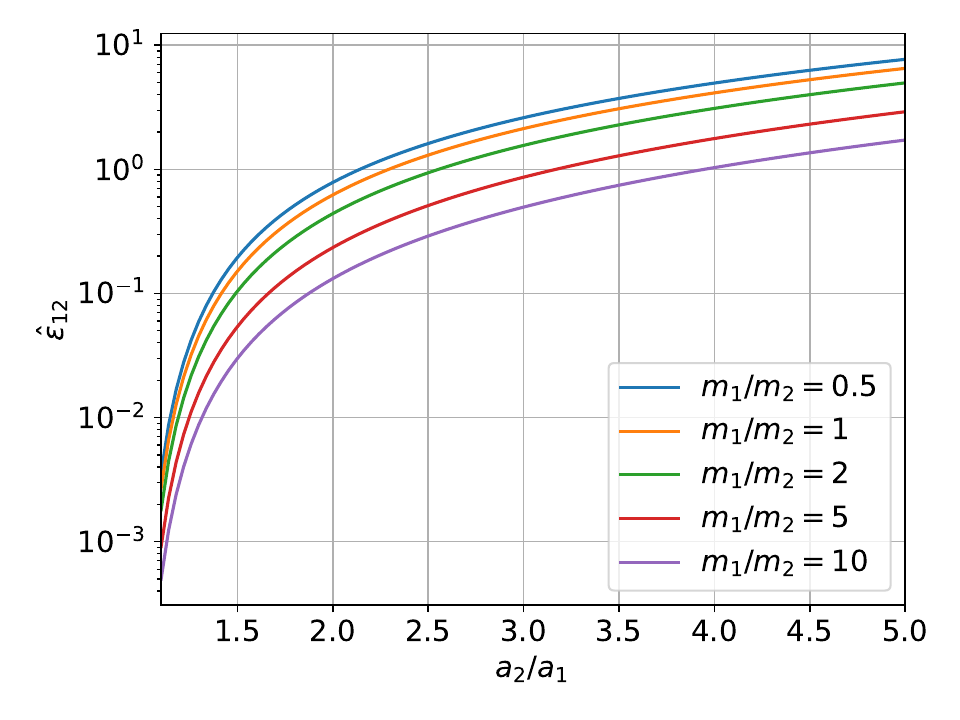}
	\caption{The rescaled coupling parameter $\hat{\varepsilon}_{12}$ as a function of the semimajor axis ratio $a_2/a_1$ of the SE system, for different mass ratios $m_1/m_2$ (see equation~\ref{eq:eps12hat}). Systems with higher $\hat{\varepsilon}_{12}$ are more easily perturbed by the outer companion.}\label{fig:eps12}
\end{figure}

The forced mutual inclination $\Delta I$ of the SEs depends on $I_{\rm p}$, $m_{\rm p}$ and $a_{\rm p}$, as well as on mutual coupling between the inner planets. Denote $\omega_{ik}$ the characteristic nodal precession frequency of planet $i$ induced by the gravitational torque from planet $k$. The frequencies of mutual interactions in the inner two planets are
\begin{equation}
	\omega_{12} = \frac{\omega_{21} L_2}{L_1} =  \frac{1}{4} n_1 \frac{m_2}{M} \left(\frac{a_1}{a_2}\right)^2 \,b_{\frac{3}{2}}^{(1)}\!\!\left(\frac{a_1}{a_2}\right),
\end{equation}
while the precession frequencies of the SEs induced by the companion are 
\begin{align}
	&\omega_{1{\rm p}} \equiv \frac{1}{4} n_1 \frac{m_{\rm p}}{M} \left(\frac{a_1}{a_{\rm p}}\right)^2\, b_{\frac{3}{2}}^{(1)}\!\!\left(\frac{a_1}{a_{\rm p}}\right),\label{eq:w1p}\\
	&\omega_{2{\rm {\rm p}}}\equiv \frac{1}{4} n_2 \frac{m_{\rm p}}{M}  \left(
	\frac{a_2}{a_{\rm p}}\right)^2 \,b_{\frac{3}{2}}^{(1)}\!\!\left(\frac{a_2}{a_{\rm p}}\right),\label{eq:w2p}
\end{align}
[see Eq. (7.11) in \cite{murraySolarSystemDynamics2000}]. Here $n_i = \sqrt{GM/a_i^3}$ and $L_i = m_i \sqrt{GMa_i}$ are the mean motion and the orbital angular momentum of planet $i$, and $b_s^{(j)}(\alpha)$ is the Laplace coefficient:
\begin{equation}
	b_s^{(j)}(\alpha) = \frac{1}{\pi}\int_{0}^{2\pi} \frac{\cos(jx)\mathrm{d}x}{(1-2\alpha\cos x +\alpha^2)^s},
\end{equation}
with $b_{\frac{3}{2}}^{(1)}(\alpha) \simeq 3\alpha (1+15\alpha^2/8+175\alpha^4/64 + ...)$ for $\alpha \ll 1$. When the two SEs are away from any mean-motion resonances \citep[see][]{rodetInclinationDynamicsResonant2021}, the forced misalignment $\Delta I$ between them induced by the inclined companion $m_{\rm p}$ depends on the coupling parameter $\varepsilon_{\rm 12, p}$ \citep{laiHidingPlanetsBig2017}, given by
\begin{equation}
	\varepsilon_{\rm 12,p} = \frac{\omega_{2p}-\omega_{1p}}{\omega_{12}+\omega_{21}} \equiv {\varepsilon}_{12} \frac{\tilde{m}_{\rm p}}{\tilde{a}_{\rm p}^3}, \label{eq:eps12p}
\end{equation}
where we have defined $\tilde{m}_{\rm p} \equiv m_{\rm p}/(1~\mathrm{M_{J}})$ and $\tilde{a}_{\rm p} = a_{\rm p} / (1~\mathrm{au})$. When $a_1, a_2 \ll a_{\rm p}$ and $m_{\rm p} \ll M$, $\varepsilon_{12}$ is independent of $a_{\rm p}$ and $m_{\rm p}$:
\begin{equation}
	\varepsilon_{12} \simeq \frac{1~\mathrm{M_J}}{m_2} \left(\frac{a_2}{1~\mathrm{au}}\right)^3 \hat{\varepsilon}_{12} \label{eq:eps12},
\end{equation}
where
\begin{equation}
	\hat{\varepsilon}_{12} =  \frac{3 a_1/a_2}{b_{{3}/{2}}^{(1)}\left({a_1}/{a_2}\right)} \frac{\left({a_2}/{a_1}\right)^\frac{3}{2}-1}{1 +{L_1}/{L_2}} \label{eq:eps12hat}.
\end{equation}
The accuracy of equation~\eqref{eq:eps12} is better than 10\% for $a_{\rm p} > 5 a_2$. Note that $\hat{\varepsilon}_{12}$ does not depend on the semimajor axis and mass scales of the SE system, and depends only on $a_2/a_1$ and $m_1/m_2$. The variation of $\hat{\varepsilon}_{12}$ with $a_2/a_1$ is shown in Figure~\ref{fig:eps12}, for different values of $m_2/m_1$.

We can now write down the forced misalignment $\Delta I$ of the two SEs as a function of the companion's inclination $I_{\rm p}$ and the coupling parameter $\varepsilon_{\rm 12,p}$. We identify two regimes: strong coupling, where $\varepsilon_{\rm 12,p} \lesssim 1$, and weak coupling, where $\varepsilon_{\rm 12,p} \gtrsim 1$. In the strong coupling regime, the companion's influence is weaker than the coupling between the SEs, and the forced inclination is proportional to $\varepsilon_{\rm 12,p}$:
\begin{equation}
	\Delta I \simeq \frac{1}{2} \sin(2I_{\rm p}) \varepsilon_{\rm 12, p},\quad (\text{for }\varepsilon_{\rm 12,p} \lesssim 1). \label{eq:Di_loweps12}
\end{equation}
On the other hand, in the weak coupling regime, the coupling between the SEs is negligible and the forced inclination reaches
\begin{equation}
	\Delta I \simeq I_{\rm p},\quad (\text{for }\varepsilon_{\rm 12,p} \gtrsim 1). \label{eq:Di}
\end{equation}
In the intermediate regime, $\Delta I$ transitions smoothly from equation~\eqref{eq:Di_loweps12} to equation~\eqref{eq:Di}. There is an exception: when $m_1 \ll m_2$, the misalignment between the SEs can become very high (larger than $I_{\rm p}$) as $\varepsilon_{\rm 12,p}$ approaches $1$ \citep{laiHidingPlanetsBig2017}. In this paper, we ignore this resonant case and assume $m_1 \gtrsim m_2$, so that equations~\eqref{eq:Di_loweps12}--\eqref{eq:Di} provide adequate description for the forced misalignment.

\begin{figure}
	\centering
	\includegraphics[width=\linewidth]{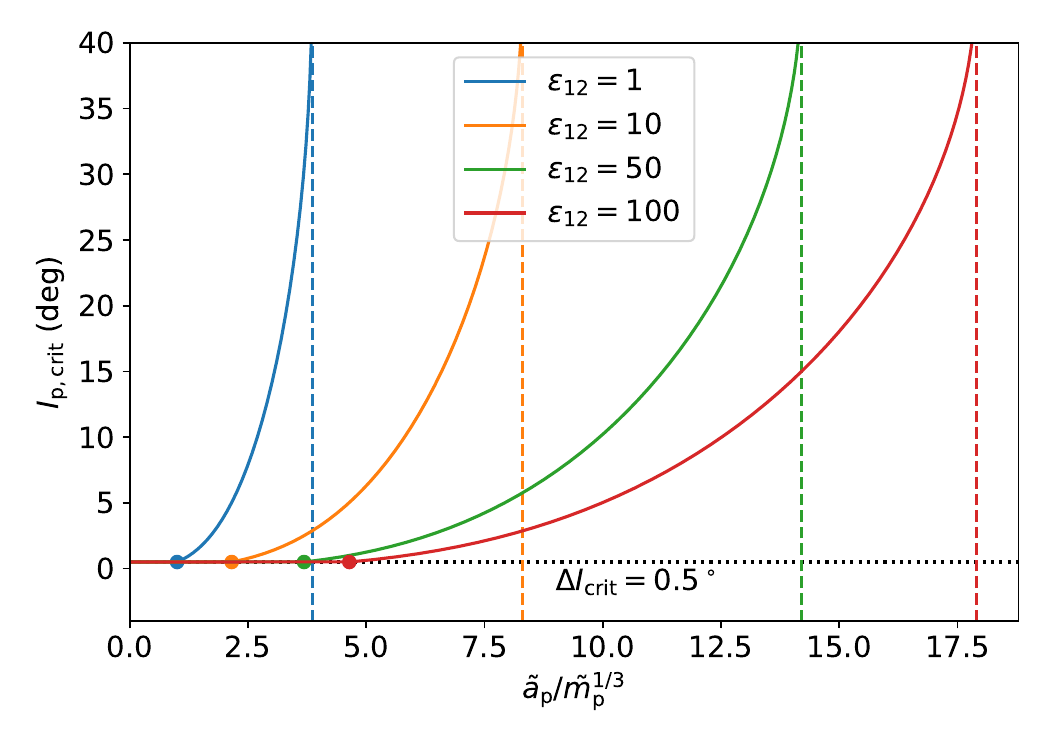}
	\caption{Minimum companion's inclination $I_{\rm p, crit}$ required to produce a forced misalignment $\Delta I_\mathrm{crit} = 0.5^\circ$ between two SEs (equations~\ref{eq:ipcrit},\ref{eq:ipcrit_loweps12}), as a function of the scaled companion semi-major axis, $\tilde{a}_{\rm p}/\tilde{m}_{\rm p}^{1/3} = a_{\rm p}/(1~\mathrm{au}) ~ (m_{\rm p}/\mathrm{M_{J}})^{-1/3}$, for different inner SEs systems, characterized by different values of $\varepsilon_{12}$ (see equation~\ref{eq:eps12}). The horizontal dashed line indicates $\Delta I_\mathrm{crit} = 0.5^\circ$, the filled circles indicate the change of regimes (between $\varepsilon_{\rm 12, p}>1$ and $\varepsilon_{\rm 12, p}<1$), while the vertical dashed lines indicate the maximum  $\tilde{a}_{\rm p}/\tilde{m}_{\rm p}^{1/3}$, beyond which it is impossible to generate $\Delta I_\mathrm{crit} = 0.5^\circ$ regardless of the $I_{\rm p}$ value (see equation~\ref{eq:amax}).}\label{fig:misalignmentvsstrength}
\end{figure}

Using equations~\eqref{eq:eps12p}--\eqref{eq:Di}, we can derive $I_{\rm p, crit}$ (that gives $\Delta I = \Delta I_\mathrm{crit}$) as a function of the rescaled semi-major axis of the companion $\tilde{a}_{\rm p}/\tilde{m}_{\rm p}^{1/3}$ for a given $\varepsilon_{12}$. In the weak coupling regime,
\begin{equation}
	\varepsilon_{\rm 12, p} \gtrsim 1 \iff \tilde{a}_{\rm p}/\tilde{m}_{\rm p}^{1/3} \lesssim \varepsilon_{12}^{1/3},
\end{equation}
we have
\begin{equation}
	I_{\rm p, crit} \simeq \Delta I_{\rm crit}, \label{eq:ipcrit}
\end{equation}
i.e. when the outer companion is close to the inner SE system, the required inclination only needs to be as low as $\Delta I_\mathrm{crit}$. In the strong coupling regime,
\begin{equation}
	\varepsilon_{\rm 12, p} \lesssim 1 \iff \tilde{a}_{\rm p}/\tilde{m}_{\rm p}^{1/3} \gtrsim \varepsilon_{12}^{1/3},
\end{equation}
we have
\begin{align}
	I_{\rm p, crit} \simeq{}& \frac{1}{2}\arcsin\left(\frac{2\Delta I_{\rm crit}}{\varepsilon_{\rm 12, p}}\right)\nonumber\\
	=&  \frac{1}{2}\arcsin\left(\frac{2\Delta I_{\rm crit} \tilde{a}_{\rm p}^3}{\varepsilon_{12} \tilde{m}_{\rm p}}\right),\label{eq:ipcrit_loweps12}
\end{align}
i.e. when the outer companion is far from the SEs, the required inclination increases with $a_{\rm p}$. The argument of the arcsin cannot be more than 1; it follows that for given $\varepsilon_{12}$ and $\tilde{m}_{\rm p}$, there is a maximum possible $\tilde{a}_{\rm p}$ for the companion, above which it cannot possibly induce a misalignment $\Delta I_\mathrm{crit}$, regardless of the value of $I_{\rm p}$. This maximum is set by $\Delta I_\mathrm{crit} = \varepsilon_{\rm 12, p}/2$, giving
\begin{align}
	\tilde{a}_{\rm p, max} ={}& \left(\frac{\varepsilon_{\rm 12} \tilde{m}_{\rm p}}{2\Delta I_\mathrm{crit}}\right)^\frac{1}{3}\nonumber\\
	\simeq{}& 3.9~\varepsilon_{12}^\frac{1}{3} \tilde{m}_{\rm p}^\frac{1}{3} \left(\frac{\Delta I_\mathrm{crit}}{0.5^\circ}\right)^{-\frac{1}{3}}. \label{eq:amax}
\end{align}
Figure~\ref{fig:misalignmentvsstrength} shows the critical $I_{\rm p}$ required to produce $\Delta I_\mathrm{crit} = 0.5^\circ$ as a function of $\tilde{a}_{\rm p}/\tilde{m}_{\rm p}^{1/3}$, for different SE systems (characterized by different values of $\varepsilon_{12}$).
In the next section, we will derive the likelihood for a flyby to excite the companion's inclination from $0$ to $I_{\rm p} > I_{\rm p, crit}$.

\section{Effect of flyby on the outer planet}
\label{sec:flyby}

\begin{figure*}
	\centering
	\includegraphics[width=\linewidth]{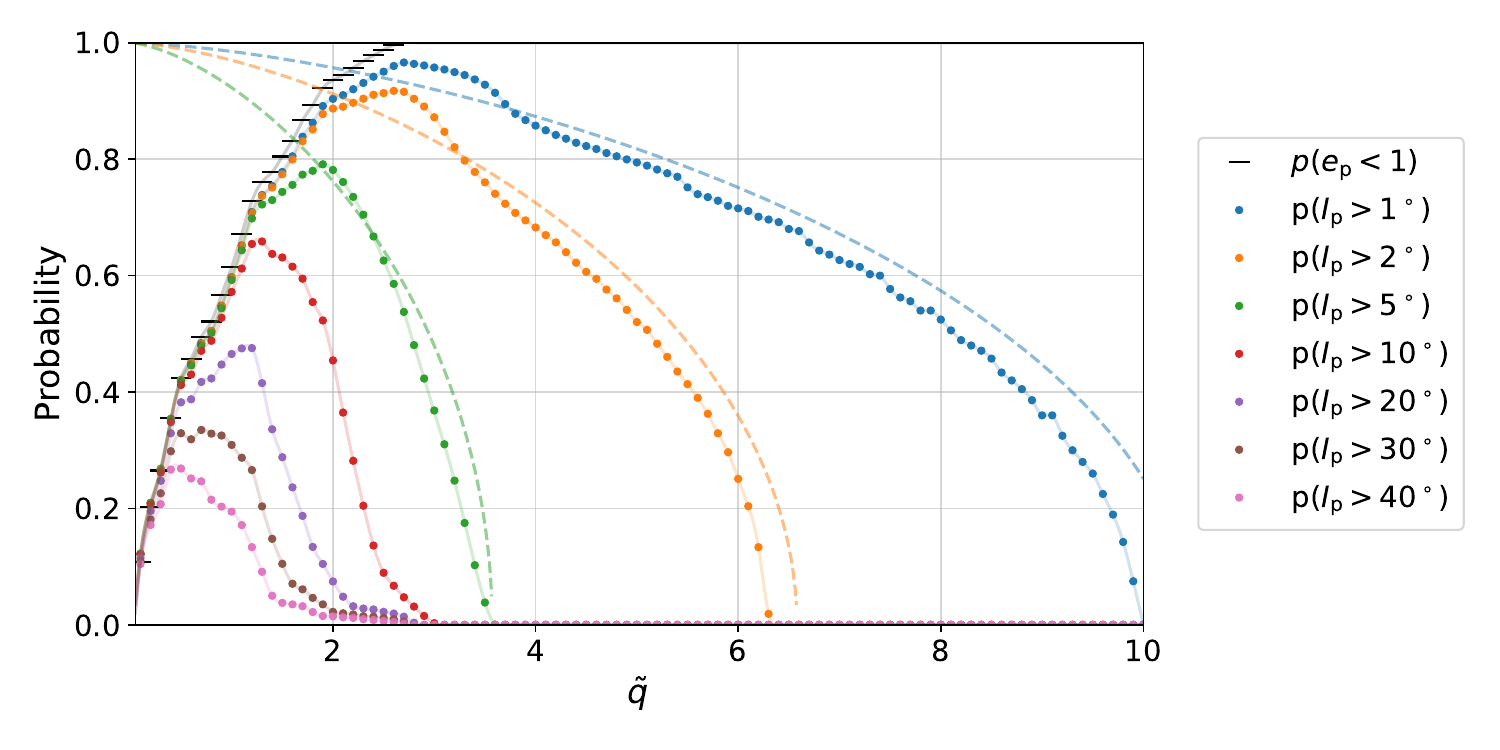}
	\caption{The probability $p(I_{\rm p} > I_{\rm p, min})$ for a flyby to raise the inclination of the outer planet to $I_{\rm p} > I_{\rm p, min}$ (while keeping the planet bound), as a function of the dimensionless distance at closest approach $\tilde{q} = q/a_{\rm p}$. Each point is computed from $30\times 20 \times 20$ N-body simulations that sample the flyby geometry. The dashed lines are the analytical results in the secular approximation, assuming a parabolic encounter (see Appendix for derivation and comparison). The difference between the approximation and the simulations is due in part to the N-body setup, where the eccentricity of the flyby is $1.1$ rather than 1. The short bars give $p(e_{\rm p} > 1)$, the probability that the planet remains bound after the flyby. }\label{fig:proba_q_inc}
\end{figure*}

In this section, we calculate the expected effect of a stellar flyby on the orbit of the outer planet/companion, and estimate the likelihood that the companion can attain $I_{\rm p} > I_{\rm p, crit}$. To this end, we perform a suite of $N$-body simulations to determine the distribution of the post-flyby inclination $I_{\rm p}$ and eccentricity $e_{\rm p}$. Following a similar approach to \cite{rodetCorrelationHotJupiters2021}, we suppose that the companion (with $m_{\rm p} \ll M_2$, the mass of the flyby star) is on an initially circular orbit, and we ignore the inner planets (their proximity to the host star effectively shields them from stellar encounter). We integrate a nearly-parabolic ($e = 1.1$) encounter between two equal-mass stars using \textsc{IAS15} from the \textsc{Rebound} package. The integration time is chosen so that the distance between $M_1$ and $M_2$ is equal to $100a_\mathrm{p}$ at the beginning and end of each simulation, and the time-step is adaptive. We sample uniformly the dimensionless distance at closest approach $\tilde{q} \equiv q/a_{\rm p}$ (where $q$ is the periastron of the encounter) from 0.05 to 10, the cosine of the inclination $\cos i$, and the argument of periastron $\omega$ of the flyby, and the initial phase $\lambda_{\rm p}$ of the planet $m_{\rm p}$. For each $\tilde{q}$, we carry out $30\times 20 \times 20$ (in $\cos i, \omega, \lambda_{\rm p}$)$= 12,000$ simulations, and obtain the final orbital elements of the planet. The effect of the flyby on $I_{\rm p}$ is mainly determined by the flyby inclination $i$, so we interpolate our data using the 30 tested inclinations to produce a denser grid of results (see Fig.~\ref{fig:ip}) and avoid under-sampling effects in computing the $I_{\rm p}$ distribution (see below).

\subsection{Effect on the inclination}

For a given $\tilde{q}$, we derive $p(e_{\rm p} < 1)$, the probability that the planet remains bound, and $p(I_{\rm p} > I_{\rm p, min})$, the probability that the planet remains bound and has a final inclination larger than some specified $I_{\rm p, min}$. These probabilities are shown in Figure~\ref{fig:proba_q_inc} as a function of $\tilde{q}$, for different values of $I_{\rm p, min}$. For $\tilde{q}\gtrsim 3$, $p(I_{\rm p} > I_{\rm p, min})$ can be derived analytically using the secular approach. This derivation is detailed in the Appendix, as well as the comparison with the $N$-body results.

\begin{figure}
	\centering
	\includegraphics[width=\linewidth]{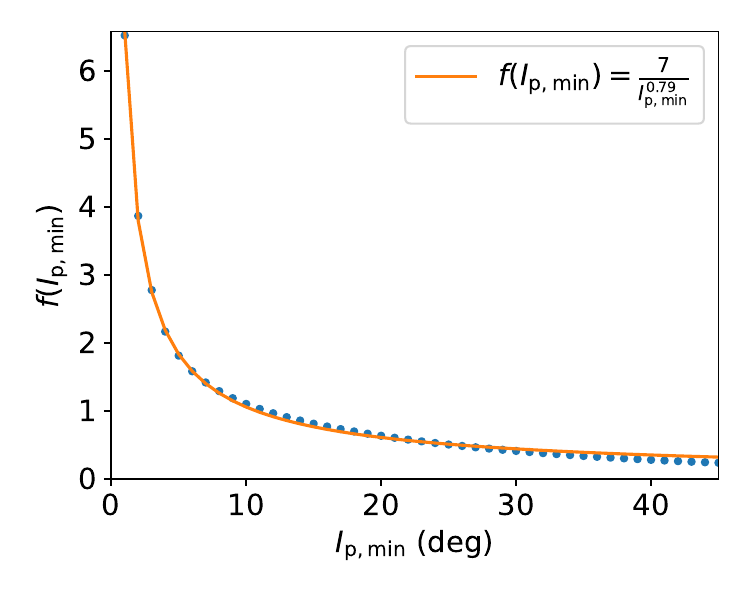}
	\caption{Integrated probability $f(I_{\rm p, min})$ (equation~\ref{eq:f}) for stellar flybys to raise the inclination of the outer planet from $0$ to $I_{\rm p} > I_{\rm p, min}$. The blue dots are computed numerically from Fig.~\ref{fig:proba_q_inc}, and the orange line is a simple power-law fit given by equation~\eqref{eq:f_fit}.}\label{fig:proba_inc}
\end{figure}

Using our result for $p(I_{\rm p} > I_{\rm p, min})$, we can derive the number of flybys that successfully raise the inclination of the outer planet (at $a_{\rm p}$) above a certain $I_{\rm p, min}$. This number can be written as a product of a scaling factor and a geometric function \citep{rodetCorrelationHotJupiters2021}:
\begin{equation}
	\mathcal{N}(I_{\rm p} > I_{\rm p, min}, a_{\rm p}) = \mathcal{N}_\mathrm{close}(a_{\rm p}) \times f(I_{\rm p, min}), \label{eq:Nflyby}
\end{equation}
where $\mathcal{N}_\mathrm{close}(a_{\rm p})$ is the number of fly-by with $q < a_{\rm p}$. Assuming that gravitational focusing is dominant in the close encounter and that the velocities of stars in the cluster follow the Maxwell--Boltzmann distribution, we have \citep{rodetCorrelationHotJupiters2021}
\begin{align}
	\mathcal{N}_\mathrm{close}(a_{\rm p}) ={}& \frac{2 \sqrt{2\pi} a_{\rm p} G M_\mathrm{tot} n_\star}{\sigma_\star}t_\mathrm{cluster}\nonumber\\
	={}&0.2 \left(\frac{t_\mathrm{cluster}}{20~\mathrm{Myr}}\right) \left(\frac{a_\mathrm{p}}{50~\mathrm{au}}\right)
	\left(\frac{M_\mathrm{tot}}{2~\mathrm{M_{\sun}}}\right)\nonumber\\ &\times\left(\frac{n_\star}{10^3~\mathrm{pc}^{-3}}\right) \left(\frac{\sigma_\star}{1~\mathrm{km/s}}\right)^{-1}, \label{eq:Nflyby_AN}
\end{align}
where $M_\mathrm{tot} = M_1+M_2$, $n_\star$ is the local stellar density of the cluster, $\sigma_\star$ its velocity dispersion, and $t_\mathrm{cluster}$ its lifetime.
In equation~\eqref{eq:Nflyby}, $f(I_{\rm p, min})$ measures the overall fraction (for all $\tilde{q}$ values) of close flybys that raise the inclination of the outer planet by $I_{\rm p} > I_{\rm p, min}$:
\begin{equation}
	f(I_{\rm p, min}) = \int_{0}^{+\infty} p(I_{\rm p}>I_{\rm p, min}) d\tilde{q}.\label{eq:f}
\end{equation}
Note that $f(I_{\rm p, min})$ can be more than 1, if the requirement $I_{\rm p}>I_{\rm p, min}$ is fulfilled for a non-negligible proportion of flybys with $q$ larger than $a_{\rm p}$.  Figure~\ref{fig:proba_inc} shows $f(I_{\rm p, min})$, obtained using the numerical results presented in Figure~\ref{fig:proba_q_inc}. The function can be approximated by a power-law fit,
\begin{equation}
	f(I_{\rm p, min}) \simeq \frac{7}{(I_{\rm p, min}/\mathrm{deg})^{0.79}}. \label{eq:f_fit}
\end{equation}
We emphasize that the function $f(I_{\rm p, min})$ is universal, in the sense that it describes the effect of a general parabolic encounter between two equal-mass stars, and can be applied to all $m_{\rm p}$ and $a_{\rm p}$ values (assuming $m_{\rm p} \ll M$). In Section~\ref{sec:rate}, we will apply equations~\eqref{eq:Nflyby} and \eqref{eq:f_fit} to ``SEs + companion'' systems to estimate the likelihood that stellar flybys can destroy the co-transiting geometry of SEs.

\subsection{Effect on the eccentricity}

\begin{figure*}
	\centering
	\includegraphics[width=\linewidth]{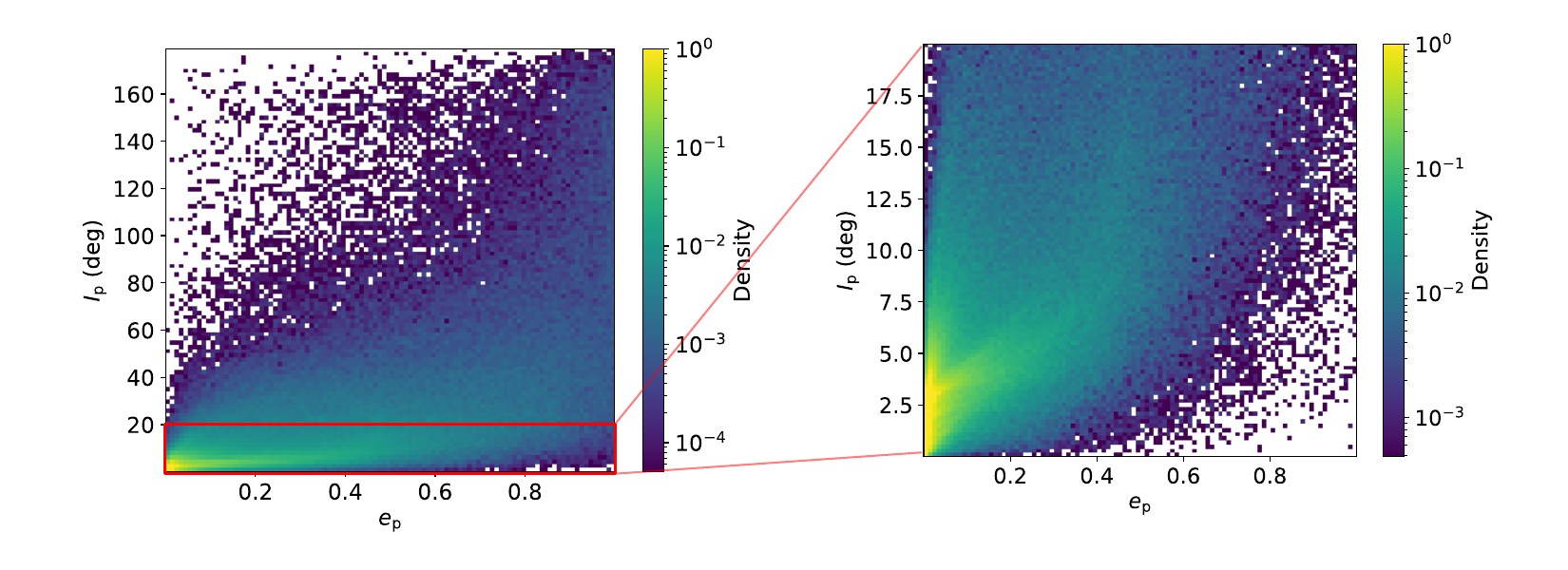}
	\includegraphics[width=\linewidth]{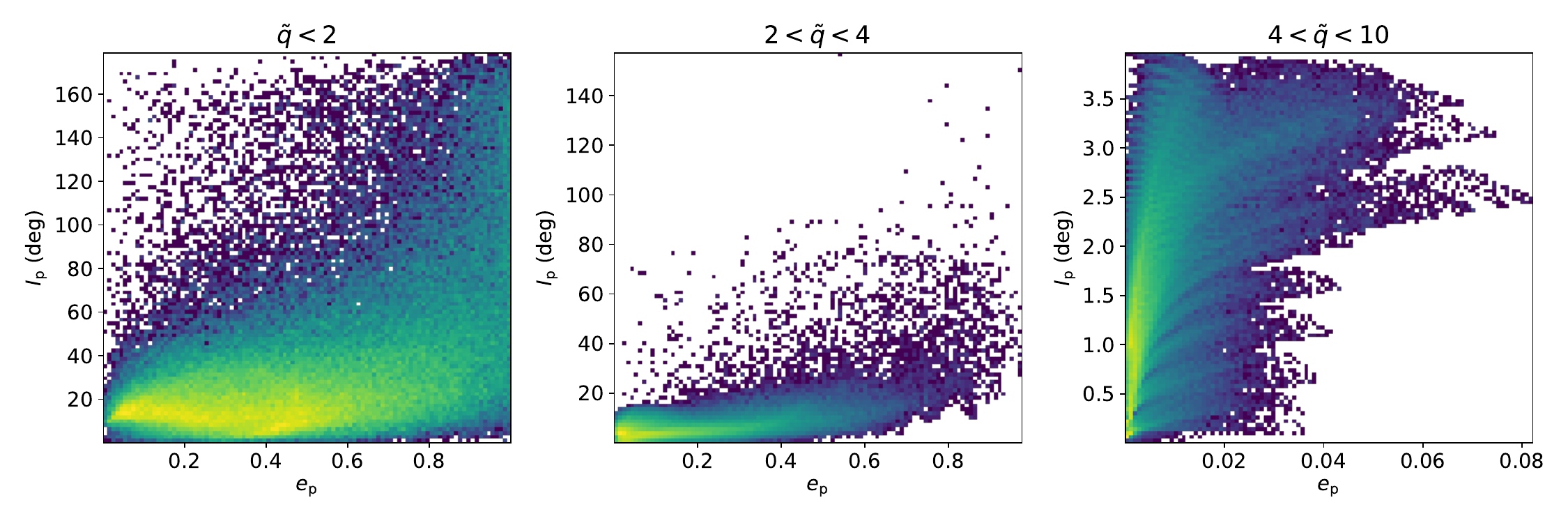}
	\caption{Correlation between the eccentricity $e_{\rm p}$ and inclination $I_{\rm p}$ of the outer planet after stellar flybys. The plots are generated from $N$-body simulations with different flyby geometries and distance at closest approaches, and the color scales logarithmically with the local density of simulations. The plots only show simulations where the planet remains bound. The upper panels gather all flyby distances at closest approach $\tilde{q}$ (with the right panel showing a zoom-in on the low inclination part of the left panel), the lower panels distinguish three different ranges of $\tilde{q}$. The stripes-like structure in the lower right panel is not physical, and is due to the discrete inclination sampling.}\label{fig:correlation_inc_ecc}
\end{figure*}

Stellar flybys not only change the inclination of the outer planet, but also excite the planet's eccentricity. Figure~\ref{fig:correlation_inc_ecc} shows the joint distribution of $I_{\rm p}$ and $e_{\rm p}$ after stellar flybys. The inclination increases are correlated to eccentricity excitations, so that the post-flyby orbit of the planet is not circular. However, its eccentricity will very likely stay low ($e_{\rm p} \lesssim 0.5$) if the inclination gain is small ($I_{\rm p} \lesssim 5^\circ$). In this case, the inclination dynamics presented in Section~\ref{sec:coupling} is largely unchanged compared to the circular orbit case. When we consider higher $I_{\rm p}$, then the eccentricity can become larger. In that case, the stability of the system is not guaranteed, so the inclination dynamics for circular orbit may not hold. In the following section, we will show that the occurrence rate of our flyby-induced misalignment scenario depends mostly on the statistics for small $I_{\rm p}$, so that we can safely ignore the eccentricity's effect on the inclination dynamics. In particular, we can neglect the cases where the flyby induces strong scatterings between planets.

\section{Occurrence rate of Super-Earth Misalignment induced by Stellar Flybys: Single Outer Planet}
\label{sec:rate}

We now compute the expected number of stellar flybys that generate a misalignment of $\Delta I > \Delta I_{\rm crit}$ between the SEs as a function of the outer planet's semimajor axis $a_{\rm p}$ and mass $m_{\rm p}$, for given SE and cluster properties. In Section~\ref{sec:coupling}, we derived the required $I_{\rm p, crit}(a_{\rm p})$ (see equations~\ref{eq:ipcrit}~\&~\ref{eq:ipcrit_loweps12}) to induce a misalignment $\Delta I_{\rm crit}$ given an outer planet at $a_{\rm p}$. From equation~\eqref{eq:Nflyby}, the number of flybys that produce $\Delta I >\Delta I_{\rm crit}$ is then
\begin{align}
	\mathcal{N}(\Delta I > \Delta I_{\rm crit}, a_{\rm p}) &=  \mathcal{N}(I_{\rm p} > I_{\rm p, crit}(a_{\rm p}), a_{\rm p})\nonumber\\
	&= \mathcal{N}_\mathrm{close}(a_{\rm p}) \times f\left(I_{\rm p, min} = I_{\rm p, crit}(a_{\rm p})\right). \label{eq:regime}
\end{align}
We previously identified two regimes for $I_{\rm p, crit}(a_{\rm p})$ (equations~\ref{eq:ipcrit}~\&~\ref{eq:ipcrit_loweps12}), depending of the coupling parameter $\varepsilon_{\rm 12, p}$. These regimes impact how $\mathcal{N}$ scales with the semimajor axis $a_{\rm p}$. In the weak coupling regime, $\varepsilon_{\rm 12, p} \gtrsim 1$ or $\tilde{a}_{\rm p} \lesssim (\varepsilon_{12} \tilde{m}_{\rm p})^{1/3}$, we have $I_{\rm p, crit}(a_{\rm p}) = \Delta I_{\rm crit}$ and thus
\begin{align}
	\mathcal{N}(\Delta I > \Delta I_{\rm crit}, a_{\rm p}) &=  \mathcal{N}_\mathrm{close}(a_{\rm p}) \times f(I_{\rm p, min} =\Delta I_{\rm crit})\nonumber \\
	&\propto t_\mathrm{cluster} n_\star \sigma_\star^{-1} M_\mathrm{tot}a_{\rm p}\left(\Delta I_{\rm crit}\right)^{-0.79}. \label{eq:regime1} 
\end{align}
In the strong coupling regime, $\varepsilon_{\rm 12, p} \lesssim 1$ or $\tilde{a}_{\rm p} \gtrsim (\varepsilon_{12} \tilde{m}_{\rm p})^{1/3}$, $I_{\rm p, crit}(a_{\rm p})$ is given by equation~\eqref{eq:ipcrit_loweps12}, and assuming $I_{\rm p, crit}(a_{\rm p}) \ll 1$ we have
\begin{equation}
	\mathcal{N}(\Delta I > \Delta I_{\rm crit}, a_{\rm p}) \propto t_\mathrm{cluster} n_\star \sigma_\star^{-1} M_\mathrm{tot} a_{\rm p}^{-1.37} \left(\frac{\varepsilon_{12} m_{\rm p} }{\Delta I_{\rm crit}}\right)^{0.79}.  \label{eq:regime2} 
\end{equation}

\begin{figure*}
	\centering
	\includegraphics[width=0.85\linewidth]{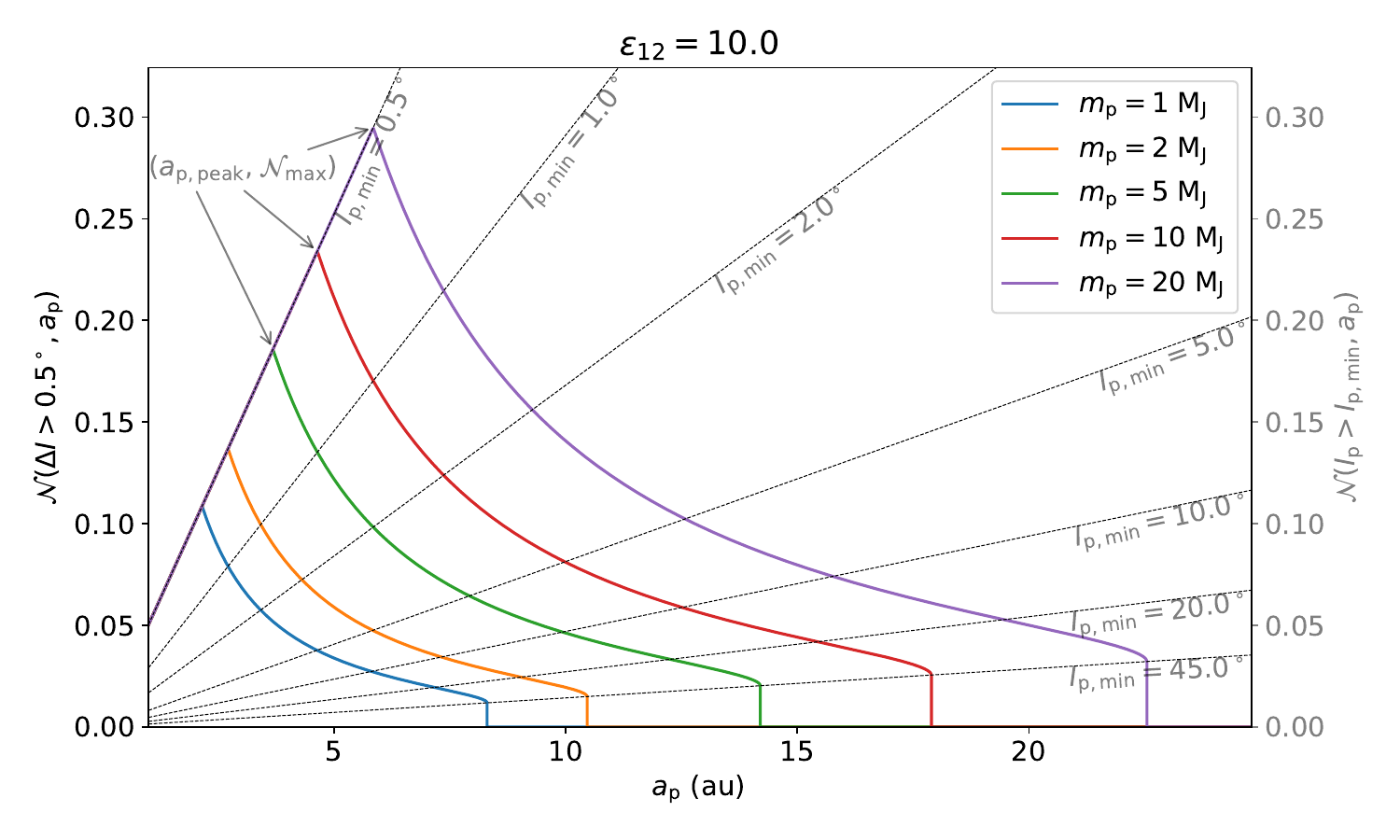}
	\caption{Number of stellar flybys that generate $\Delta I >\Delta I_{\rm crit} = 0.5^\circ$ in the SE system as a function of $a_{\rm p}$ of the outer planet companion, for different values of $m_{\rm p}$ (solid lines; see equations~\ref{eq:regime},\ref{eq:regime1},\ref{eq:regime2}). Each curve terminates at $a_{\rm p, max}$, given by equation~\eqref{eq:amax}. The peak of each curve is located at $(a_{\rm p, peak}, \mathcal{N}_{\rm max})$. The inner SE system is characterized by $\varepsilon_{12} = 10$ (equation~\ref{eq:eps12}), and the cluster parameters are fixed to the fiducial values of equation~\eqref{eq:Nflyby_AN}: $t_\mathrm{cluster} = 20$ Myr, $\sigma_\star = 1$ km.s$^{-1}$, $n_\star = 10^3$ pc$^{-3}$, and $M_\mathrm{tot} = 2 M_\odot$.The light dashed lines give $\mathcal{N}(I_{\rm p} > I_{\rm p, min}, a_{\rm p})$ for different values of $I_{\rm p, min}$ (see equation~\ref{eq:Nflyby}).}\label{fig:Nflyby_isoinc}
\end{figure*}

\begin{figure*}
	\centering
	\includegraphics[width=0.85\linewidth]{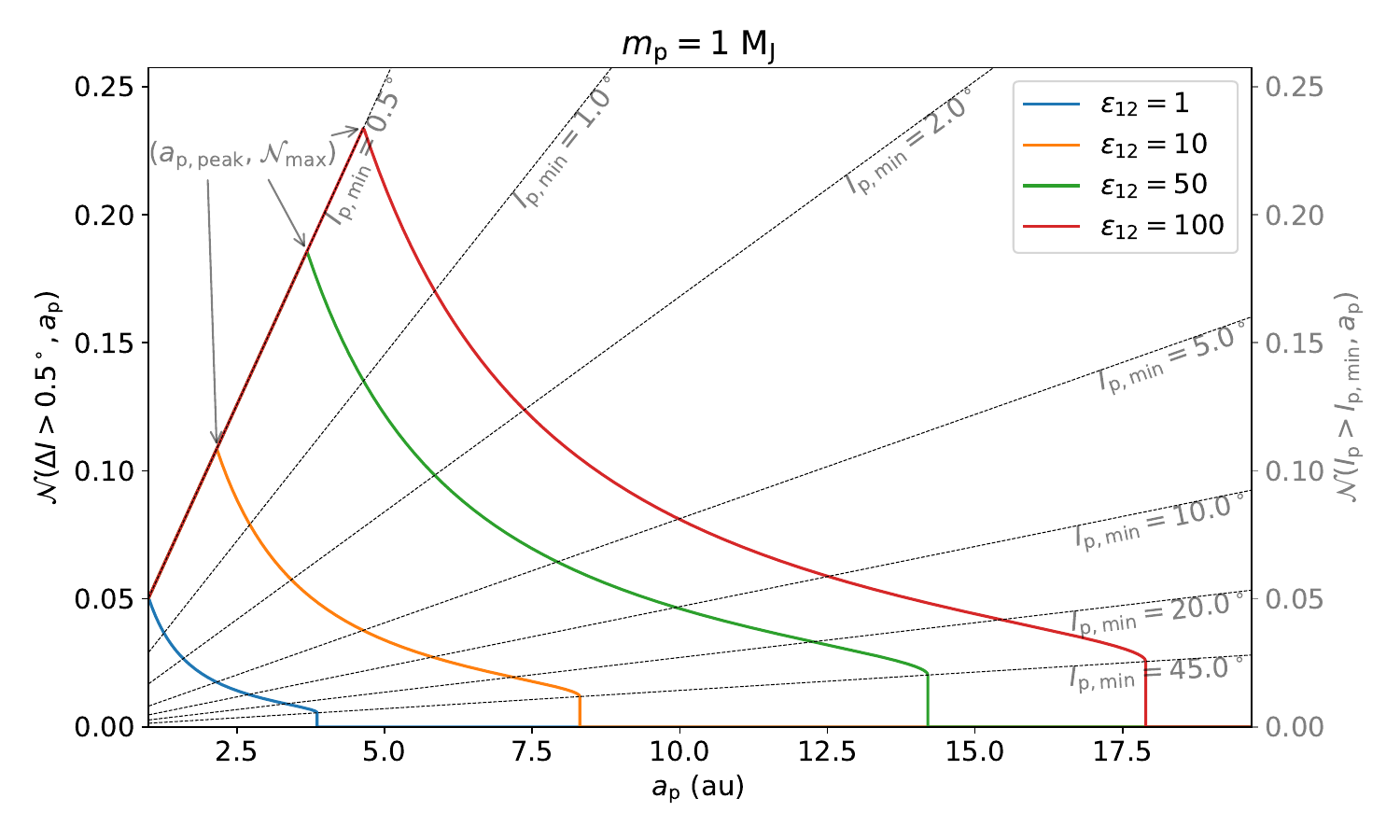}
	\caption{Same as Figure~\ref{fig:Nflyby_isoinc}, except for a fixed $m_{\rm p} = 1~\mathrm{M_J}$ and different coupling parameter $\varepsilon_{12}$ (see equation~\ref{eq:eps12}) for the inner SE system.}\label{fig:Nflyby_isoinc_variations}
\end{figure*}

Figures~\ref{fig:Nflyby_isoinc} and \ref{fig:Nflyby_isoinc_variations} show $\mathcal{N}(\Delta I > 0.5^\circ, a_{\rm p})$ as a function of the semimajor axis of the outer planet $a_{\rm p}$, for different values of $m_{\rm p}$ and $\varepsilon_{12}$ (which characterizes the coupling of the inner SE system, see equation~\ref{eq:eps12}). The two regimes (linear and decreasing power-law dependence of $a_{\rm p}$) are clearly visible on the plots. Moreover, there is a maximum $a_{\rm p, max}$ above which no $I_{\rm p}$ can induce the required $\Delta I_{\rm crit}$ (see equation~\ref{eq:amax}), and thus $\mathcal{N}(\Delta I>\Delta I_{\rm crit}, a_{\rm p}) = 0$ for  $a_{\rm p} > a_{\rm p, max}$.

\begin{figure}
	\centering
	\includegraphics[width=\linewidth]{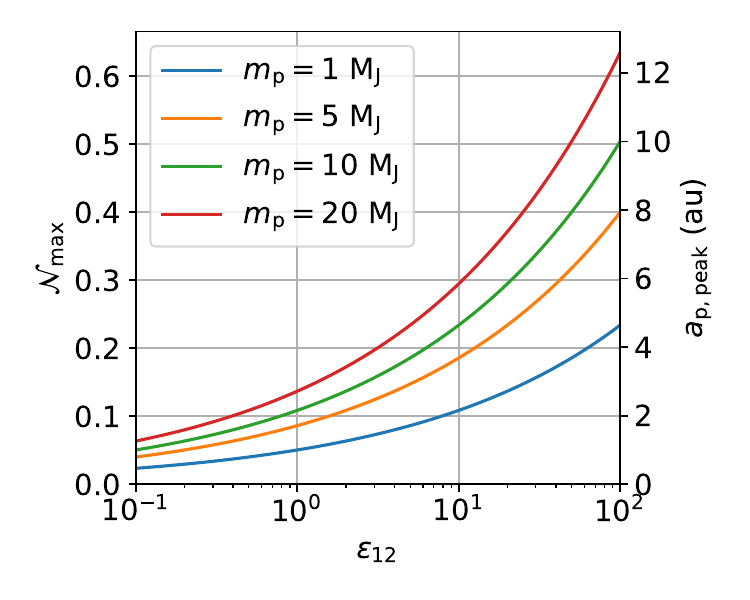}
	\caption{The maximum possible number of flybys $\mathcal{N}_{\rm max}$ that can produce $\Delta I > 0.5^\circ$ in the SE system and corresponding $a_{\rm p, peak}$ as a function of $\varepsilon_{12}$, for different $m_{\rm p}$ (see equations~\ref{eq:apnmax}--\ref{eq:nmax}). The $\mathcal{N}_{\rm max}$ value uses the cluster parameters $t_\mathrm{cluster} = 20$ Myr, $\sigma_\star = 1$ km/s and $n_\star = 10^3$ pc$^{-3}$. $\mathcal{N}_{\rm max}$ scales with $m_{\rm p}^{1/3}$, but can be easily rescaled for different parameters (see equation~\ref{eq:Nflyby_AN}) and different $m_{\rm p}$ (since $\mathcal{N}_\mathrm{max} \propto m_{\rm p}^{1/3}$).}\label{fig:Nflyby_max_vs_eps12}
\end{figure}

The maximum of each curve in Figs.~\ref{fig:Nflyby_isoinc}--\ref{fig:Nflyby_isoinc_variations}, $\mathcal{N}_{\rm max}$, is reached at $\varepsilon_{\rm 12,p} \simeq 1$, or $\tilde{a}_{\rm p} \simeq (\varepsilon_{12} \tilde{m}_{\rm p})^{1/3}$. It requires a companion inclination $I_{\rm p}$ no more than $\Delta I_\mathrm{crit}$, justifying our assumption to neglect the eccentricity effect (see Section~\ref{sec:flyby}, Fig.~\ref{fig:correlation_inc_ecc}). They are given by
\begin{align}
	a_{\rm p, peak} \simeq {}& 2.2~\mathrm{au} \left(\frac{\varepsilon_{12}}{10}\right)^\frac{1}{3} \left(\frac{m_{\rm p}}{1~\mathrm{M_{J}}}\right)^\frac{1}{3}, \label{eq:apnmax}  \\
	\mathcal{N}_{\rm max} \simeq{}& 0.1 \left(\frac{\Delta I_{\rm crit}}{0.5^\circ}\right)^{-0.79}  \left(\frac{{\varepsilon}_{12}}{10}\right)^\frac{1}{3} \left(\frac{m_{\rm p}}{1~\mathrm{M_{J}}}\right)^\frac{1}{3}  \left(\frac{t_\mathrm{cluster}}{20~\mathrm{Myr}}\right)\nonumber\\
	&\times \left(\frac{M_\mathrm{tot}}{2~\mathrm{M_{\sun}}}\right) \left(\frac{n_\star}{10^3~\mathrm{pc}^{-3}}\right) \left(\frac{\sigma_\star}{1~\mathrm{km.s^{-1}}}\right)^{-1} \label{eq:nmax}
\end{align}
We recall that $\varepsilon_{12}$ varies with $a_2^3/m_2$ for fixed $a_2/a_1$ and $m_2/m_1$ (see equation~\ref{eq:eps12}), so that both $\mathcal{N}_{\rm max}$ and $a_\mathrm{p,peak}$ are proportional to $a_2$, i.e. to the scale of the inner SE system. Figure~\ref{fig:Nflyby_max_vs_eps12} displays $\mathcal{N}_{\rm max}$ and  $a_{\rm p, peak}$ as a function of $\varepsilon_{12}$ for different $m_{\rm p}$. Note that since both $\mathcal{N}_{\rm max}$ and $a_\mathrm{p,peak}$ scale with $(\varepsilon_{12} m_{\rm p})^{1/3}$ in the same way, they can be plotted with the same curves with different vertical scales. From equation~\eqref{eq:eps12} and Fig.~\ref{fig:eps12}, we see that for typical SE systems, $\varepsilon_{12} = 2.5 (a_2/0.2~\mathrm{au})^3 (m_2/10~\mathrm{M_{\earth}})^{-1} \hat{\varepsilon}_{12}$ and $\hat{\varepsilon}_{12} \sim 1$ (for $a_2/a_1 \sim  2$). Equation~\eqref{eq:nmax} and Fig.~\ref{fig:Nflyby_max_vs_eps12} then indicate $\mathcal{N} \lesssim 0.1$ (subject to uncertainties related to various SE and cluster parameters), suggesting that the co-transiting geometry of most SE systems is not affected by stellar flybys. In the next section, we will consider how this result changes when SE systems have two external companions.

\section{Super-Earth systems with two exterior companions}
\label{sec:rate2p}

\begin{figure}
	\centering
	\includegraphics[width=\linewidth]{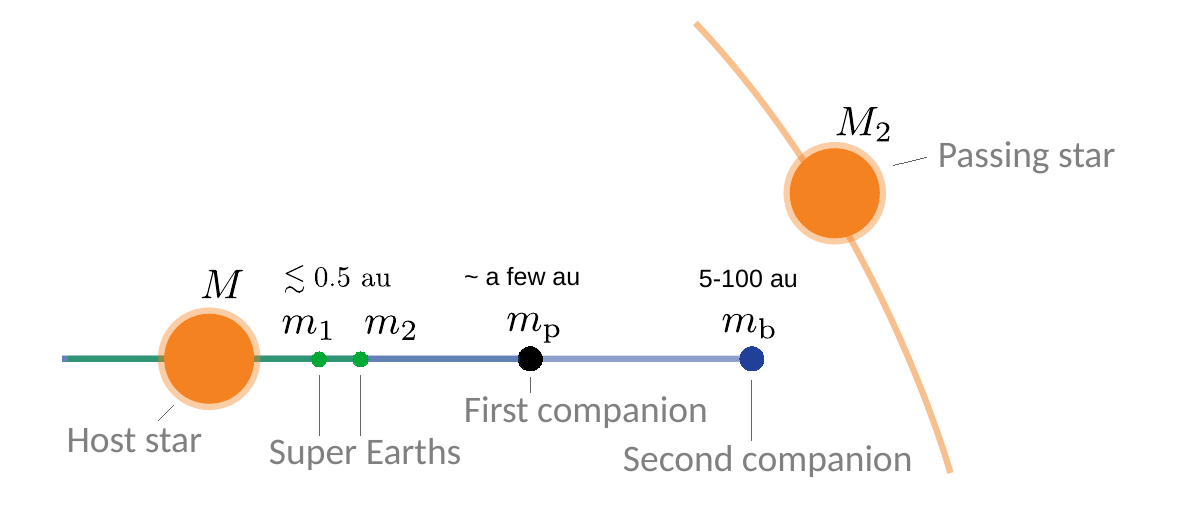}
	\caption{Schematic of the flyby-induced misalignment cascade scenario considered in Section~\ref{sec:rate2p}: an initially coplanar system comprising two inner planets (SEs) and two outer companions encounters a passing star.}\label{fig:notations2}
\end{figure}

In the previous section, we have seen that flyby-induced misalignment of SEs by way of a single outer companion/planet is somewhat inefficient (see equation~\ref{eq:nmax}). In this section we add a second companion $m_{\rm b}$ in a circular orbit (semimajor axis $a_{\rm b}$) around the host star (see Fig.~\ref{fig:notations2}). This second companion can be farther away, thus increasing the effect cross section for flybys. The flyby-induced misalignment $I_{\rm b}$ on this second companion will misalign the first companion by $I_{\rm p}$, which will then increase the mutual inclination $\Delta I$ between the SEs.

Similarly to Section~\ref{sec:rate}, we focus on the cases where the architecture of the system pre-flyby is coplanar. For companions at large separations \citep[$\gtrsim 50$ au,][]{tokovininOrbitAlignmentTriple2017}, this is not the most likely configuration. If a wide companion form with a primordial inclination, then flybys are not needed to misalign the inner system. In such systems, the stellar environment is expected to play a less important role than the initial conditions in the inner system dynamics. Here, we are interested in systems that are significantly impacted by the stellar environment, so we focus on initially coplanar companions.

\begin{figure}
	\centering
	\includegraphics[width=\linewidth]{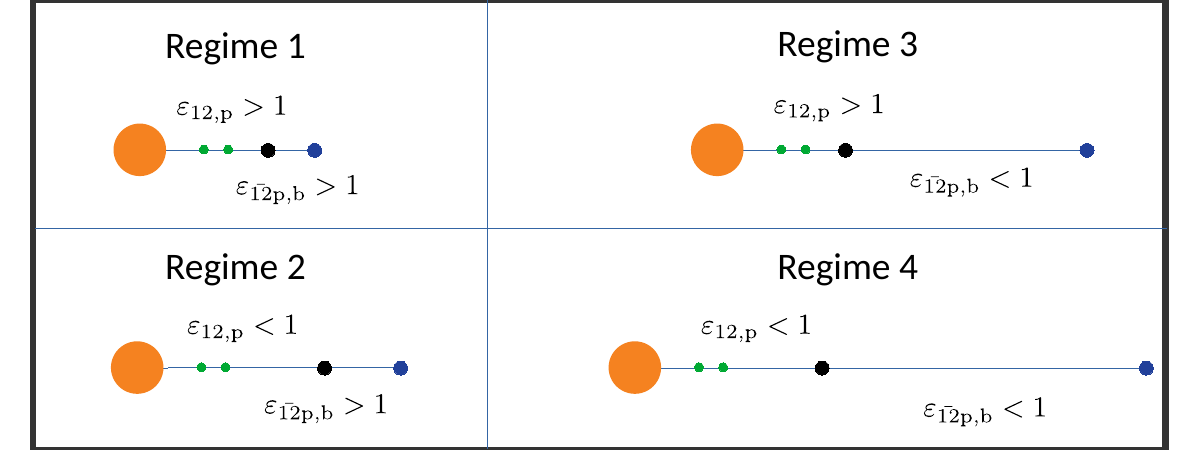}
	\caption{Schematic of the four different regimes characterized by the coupling parameters $\varepsilon_{\rm 12, p}$ (see equation~\ref{eq:eps12p}) and $\varepsilon_{\rm \bar{12}p, b}$ (see equation~\ref{eq:eps12pb}). When $\varepsilon_{\rm 12, p} \gtrsim 1$ ($\lesssim 1$), the first companion $m_{\rm p}$ has a stronger (weaker) influence on the SEs ($m_1$ and $m_2$) compared to the mutual coupling between $m_1$ and $m_2$; when $\varepsilon_{\rm \bar{12}p, b} \gtrsim 1$ ($\lesssim 1$), the second companion $m_{\rm b}$ has a stronger (weaker) influence on the inner "SEs+$m_{\rm p}$" system compared to the mutual coupling between the SEs and $m_{\rm p}$. In this schematic, a companion is strong if its distance to the inner bodies is similar to the distance between the inner bodies or if it is much more massive.}\label{fig:regimes}
\end{figure}

Equations \eqref{eq:Di_loweps12}--\eqref{eq:Di} giving the forced mutual inclination $\Delta I$ of the SEs as a function of the misalignment $I_{\rm p}$ of the first companion are still valid. However, $I_{\rm p}$ is not a direct product of the flyby anymore: it is induced by the misalignment of the second companion $m_{\rm b}$. The relationship between $I_{\rm p}$ and $I_{\rm b}$ is controlled by a new coupling parameter $\varepsilon_{\rm \bar{12}p,b}$, which characterizes the forcing strength of the second companion on the inner (SEs+$m_{\rm p}$) system compared to the mutual coupling between the SEs and $m_{\rm p}$. To evaluate this parameter, we note that when the two SEs are strongly coupled, we can treat them as a single body (denoted by $\bar{12}$), and their orbital axis $\vec{L_{\bar{12}}}=\vec{L_1}+\vec{L_2}$ precesses around $\vec{L_{\rm p}}$ with the characteristic frequency
\begin{equation}
	\omega_{\rm \bar{12}p} = \frac{L_1\omega_{1p} + L_2\omega_{2p}}{L_1+L_2},
\end{equation}
where $\omega_{1p}$ and $\omega_{2p}$ are given by equations~\eqref{eq:w1p}--\eqref{eq:w2p}. For $a_1,a_2\ll a_{\rm p}$, we can define the effective semi-major axis of $\bar{12}$ via [see Eqs~\ref{eq:w1p}--\ref{eq:w2p} with $b_{3/2}^{(1)}(\alpha) \simeq 3\alpha$]
\begin{equation}
	\omega_{\rm \bar{12}p} \simeq \frac{3}{4} n_{12} \frac{m_{\rm p}}{M} \left(\frac{a_{\bar{12}}}{a_{\rm p}}\right)^3 = \frac{3}{4} \sqrt{\frac{GM}{a_{\bar{12}}^3}} \frac{m_{\rm p}}{M} \left(\frac{a_{\bar{12}}}{a_{\rm p}}\right)^3,
\end{equation}
which gives
\begin{equation}
	a_{\bar{12}} = \left(\frac{m_1 a_1^2 + m_2 a_2^2}{m_1\sqrt{a_1} + m_2\sqrt{a_2}}\right)^\frac{2}{3}.
\end{equation}
Similarly, the characteristic precession rate of $\vec{L_{\bar{12}}}$ around $\vec{L_{\rm b}}$ is given by
\begin{equation}
	\omega_{\rm \bar{12}b} = \frac{L_1\omega_{1b} + L_2\omega_{2b}}{L_1+L_2} = \frac{3}{4} n_{12} \frac{m_{\rm b}}{M} \left(\frac{a_{\bar{12}}}{a_{\rm b}}\right)^3.
\end{equation}
Thus, analogous to equation~\eqref{eq:eps12p}, we can define the coupling parameter
\begin{equation}
	\varepsilon_{\rm \bar{12}p,b} = \frac{\omega_{\rm pb}-\omega_{\rm \bar{12}b}}{\omega_{\rm \bar{12}p}+\omega_{\rm p \bar{12}}} \simeq \frac{\omega_{\rm pb}}{\omega_{\rm \bar{12}p}} =  \frac{m_{\rm b}}{m_{\rm p}} \left(\frac{a_{\rm p}}{a_{\rm b}}\right)^3 \left(\frac{a_{\rm p}}{a_{\bar{12}}}\right)^\frac{3}{2},\label{eq:eps12pb}
\end{equation}
where in the second equality we have used $\omega_{\rm p \bar{12}} = (L_{\bar{12}}/L_{\rm p}) \omega_{\rm \bar{12}p} \ll \omega_{\rm \bar{12}p}$ and $\omega_{\rm \bar{12}b}\ll \omega_{\rm pb}$---these are valid for $a_{\bar{12}} < a_{\rm p} \ll a_{\rm b}$ and $m_1 , m_2 \ll m_{\rm p}, m_{\rm b}$. When $\varepsilon_{\rm \bar{12}p,b} \ll 1$, then the second companion's influence is weaker than the coupling between the first companion ($m_{\rm p}$) and the SEs, and the forced inclination between $m_{\rm p}$ and the SE system is
\begin{equation}
	I_{\rm p} \simeq \frac{1}{2} \sin(2I_{\rm b}) \varepsilon_{\rm \bar{12}p,b}.
\end{equation}
On the other hand, when $\varepsilon_{\rm \bar{12}p,b} \gg 1$, the coupling between $m_{\rm p}$ and the SEs is negligible and the forced inclination becomes
\begin{equation}
	I_{\rm p} \simeq I_{\rm b} 
\end{equation}
The finite $I_{\rm p}$ will then lead to misalignment $\Delta I$ between the two SEs (see Section~\ref{sec:coupling}). The critical $I_{\rm b, crit}$ to induce a misalignment $\Delta I_{\rm crit}$ between the SEs depends on the two coupling parameters, $\varepsilon_{\rm 12,p}$ and $\varepsilon_{\rm \bar{12}p,b}$. This leads to four regimes (Fig.~\ref{fig:regimes}):
\begin{enumerate}
	\item $\varepsilon_{\rm 12, p} \gtrsim 1$ and $\varepsilon_{\rm \bar{12}p, b} \gtrsim 1$\\
	The first companion $m_{\rm p}$ has a strong influence on the SEs, and the second companion $m_{\rm b}$ has a strong influence on the inner "SEs+$m_{\rm p}$" system. The misalignment $I_{\rm b}$ of $m_{\rm b}$ generated by a flyby is then directly transmitted to $m_{\rm p}$ and the SEs (i.e. $\Delta I \simeq I_{\rm p} \simeq I_{\rm b}$). 
	\item $\varepsilon_{\rm 12, p} \lesssim 1$ and $\varepsilon_{\rm \bar{12}p, b} \gtrsim 1$\\
	The first companion has a weak influence on the SEs, but the second companion has a strong influence on the inner system. The misalignment of $m_{\rm b}$ is directly transmitted to $m_{\rm p}$ (ie $I_{\rm p} \simeq I_{\rm b}$), but the mutual inclination $\Delta I$ of the SEs is reduced from $I_{\rm p}$ (see equation~\ref{eq:Di_loweps12}).
	\item $\varepsilon_{\rm 12, p} \gtrsim 1$ and $\varepsilon_{\rm \bar{12}p, b} \lesssim 1$\\
	The first companion has a strong influence on the SEs, but the second companion has a weak influence on the inner system. The misalignment $I_{\rm p}$ of $m_{\rm p}$ is then reduced from $I_{\rm b}$, while it is directly transmitted to the SEs (ie $\Delta I \simeq I_{\rm p}$).
	\item $\varepsilon_{\rm 12, p} \lesssim 1$ and $\varepsilon_{\rm \bar{12}p, b} \lesssim 1$\\
	The first companion has a weak influence on the SEs, and the second companion also has a weak influence on the inner system. The misalignment of $m_{\rm p}$ is then reduced from $I_{\rm b}$, and the mutual inclination of the SEs is also reduced from $I_{\rm p}$.
\end{enumerate}
The expression of $I_{\rm b, crit}$ in each regime is
\begin{equation}
	I_{\rm b, crit} \simeq \left\{ \begin{array}{ccc}
		\Delta I_{\rm crit} & \text{Regime 1}\\
		\frac{1}{2}\arcsin\left(\frac{2\Delta I_{\rm crit}}{\varepsilon_{\rm 12, p}}\right)& \text{Regime 2}\\
		\frac{1}{2}\arcsin\left(\frac{2\Delta I_{\rm crit}}{\varepsilon_{\rm \bar{12}p,b}}\right)& \text{Regime 3}\\
		\frac{1}{2}\arcsin\left(\frac{1}{\varepsilon_{\rm \bar{12}p, b}}\arcsin\left(\frac{2\Delta I_{\rm crit}}{\varepsilon_{\rm 12, p}}\right)\right)& \text{Regime 4}
	\end{array} \right.
\end{equation} 
The maximum possible value for the semimajor axis of the first companion derived in equation~\eqref{eq:amax} still holds,  i.e.
\begin{equation}
	\tilde{a}_{\rm p, max} = \left(\frac{\varepsilon_{\rm 12} \tilde{m}_{\rm p}}{2\Delta I_\mathrm{crit}}\right)^\frac{1}{3}.
\end{equation}
In addition, there is a maximum value of $a_{\rm b}$ beyond which $\Delta I > \Delta I_\mathrm{crit}$ can never be produced; this $a_{\rm b, max}$ is set by $I_{\rm p^, crit} = \varepsilon_{\rm \bar{12}p,b}/2$, with $I_{\rm p, crit}$ given by equations~\eqref{eq:ipcrit} and \eqref{eq:ipcrit_loweps12}, depending on whether $\varepsilon_{\rm 12, p} \gtrsim 1$ or $\varepsilon_{\rm 12, p} \lesssim 1$. Thus we have
\begin{equation}
	a_{\rm b, max} = \left\{ \begin{array}{ccc}
		a_{\rm p} \left(\frac{a_{\rm p}}{a_{\bar{12}}}\right)^\frac{1}{2} \left(\frac{m_{\rm b}}{m_{\rm p}} \frac{1}{2\Delta I_\mathrm{crit}} \right)^\frac{1}{3} & \text{if }\varepsilon_{\rm 12, p} \gtrsim 1\\
		a_{\rm p} \left(\frac{a_{\rm p}}{a_{\bar{12}}}\right)^\frac{1}{2} \left(\frac{m_{\rm b}}{m_{\rm p}} \frac{\varepsilon_{\rm 12, p}}{2\Delta I_\mathrm{crit}} \right)^\frac{1}{3} & \text{if }\varepsilon_{\rm 12, p} \lesssim 1
	\end{array} \right.\label{eq:abmax}
\end{equation} 

\begin{figure*}
	\centering
	\includegraphics[width=0.85\linewidth]{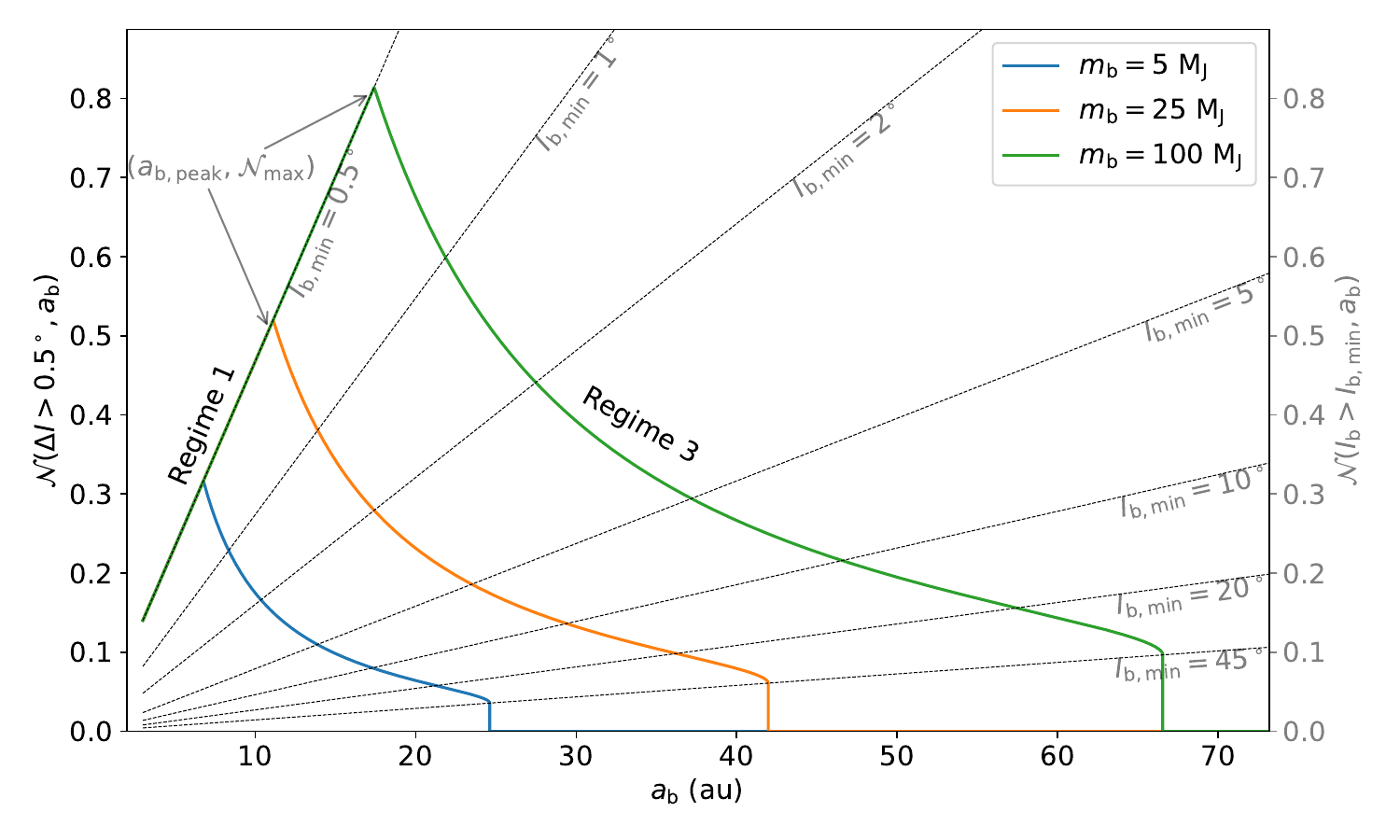}
	\caption{Number of stellar flybys that generate $\Delta I \gtrsim \Delta I_{\rm crit} = 0.5^\circ$ in the SE system as a function of $a_{\rm b}$ of the outermost companion, for different values of $m_{\rm b}$ (solid lines; see equation~\ref{eq:nflyby_twop}). Each curve terminates at $a_{\rm b, max}$, given by equation~\eqref{eq:abmax}. The inner SE system is characterized by $\varepsilon_{12} = 10$ (equation~\ref{eq:eps12}), $a_{\bar{12}} = 0.5$ au, and the inner companion's parameters are fixed to $a_{\rm p} = 2$ au and $m_{\rm p} = 1~\mathrm{M_J}$, so that $\varepsilon_{\rm 12, p} \gtrsim 1$. This situation corresponds to regimes 1 (for small $a_{\rm b}$) and 3 (for high $a_{\rm b}$). The cluster parameters are fixed to the fiducial values of equation~\eqref{eq:Nflyby_AN}: $t_\mathrm{cluster} = 20$ Myr, $\sigma_\star = 1$ km/s and $n_\star = 10^3$ pc$^{-3}$. The light dashed lines give $\mathcal{N}(I_{\rm b} > I_{\rm b, min}, a_{\rm b})$ for different values of $I_{\rm b, min}$ (see equation~\ref{eq:Nflyby}, applied to the outermost companion). See Figures~\ref{fig:Nflyby_isoinc}--\ref{fig:Nflyby_isoinc_variations} for the cases of SEs with a single companion.}\label{fig:Nflyby_isoinc_twop}
\end{figure*}

\begin{figure*}
	\centering
	\includegraphics[width=0.85\linewidth]{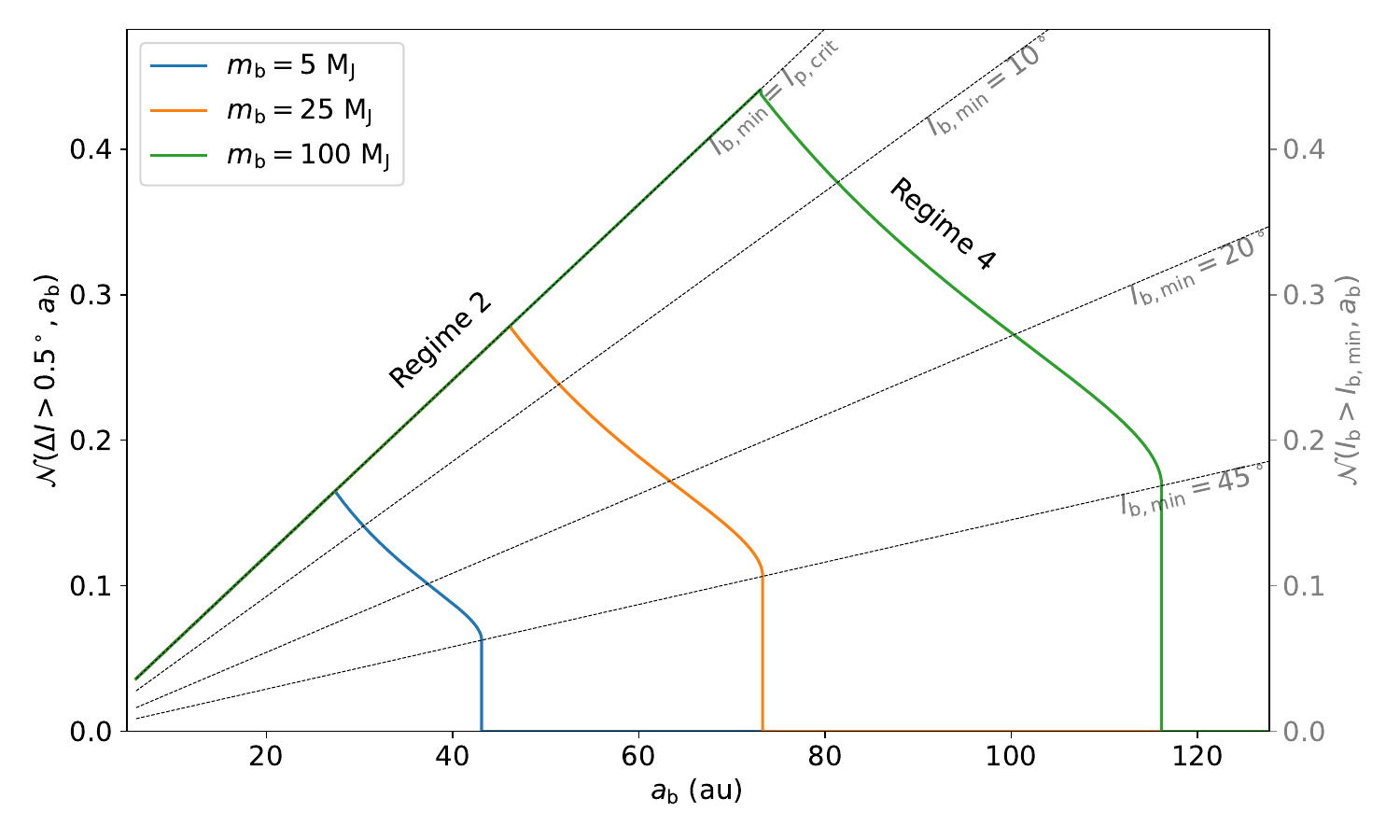}
	\caption{Same as Figure~\ref{fig:Nflyby_isoinc_twop}, except for $a_{\rm p} = 5$ au, corresponding to $\varepsilon_{\rm 12, p} = 0.07$ (see equation~\ref{eq:eps12}). For this value of the coupling parameter, the first companion $m_{\rm p}$ should have a minimum $I_{\rm p, crit} \sim 7^\circ > \Delta I_\mathrm{crit}$ (see equation~\ref{eq:ipcrit_loweps12}) to generate a misalignment $\Delta I >  \Delta I_\mathrm{crit}$ within the inner SE system. This case corresponds to regimes 2 (for small $a_{\rm b}$) and 4 (for high $a_{\rm b}$).}\label{fig:Nflyby_isoinc_twop_5au}
\end{figure*}

\begin{figure*}
	\centering
	\includegraphics[width=.8\linewidth]{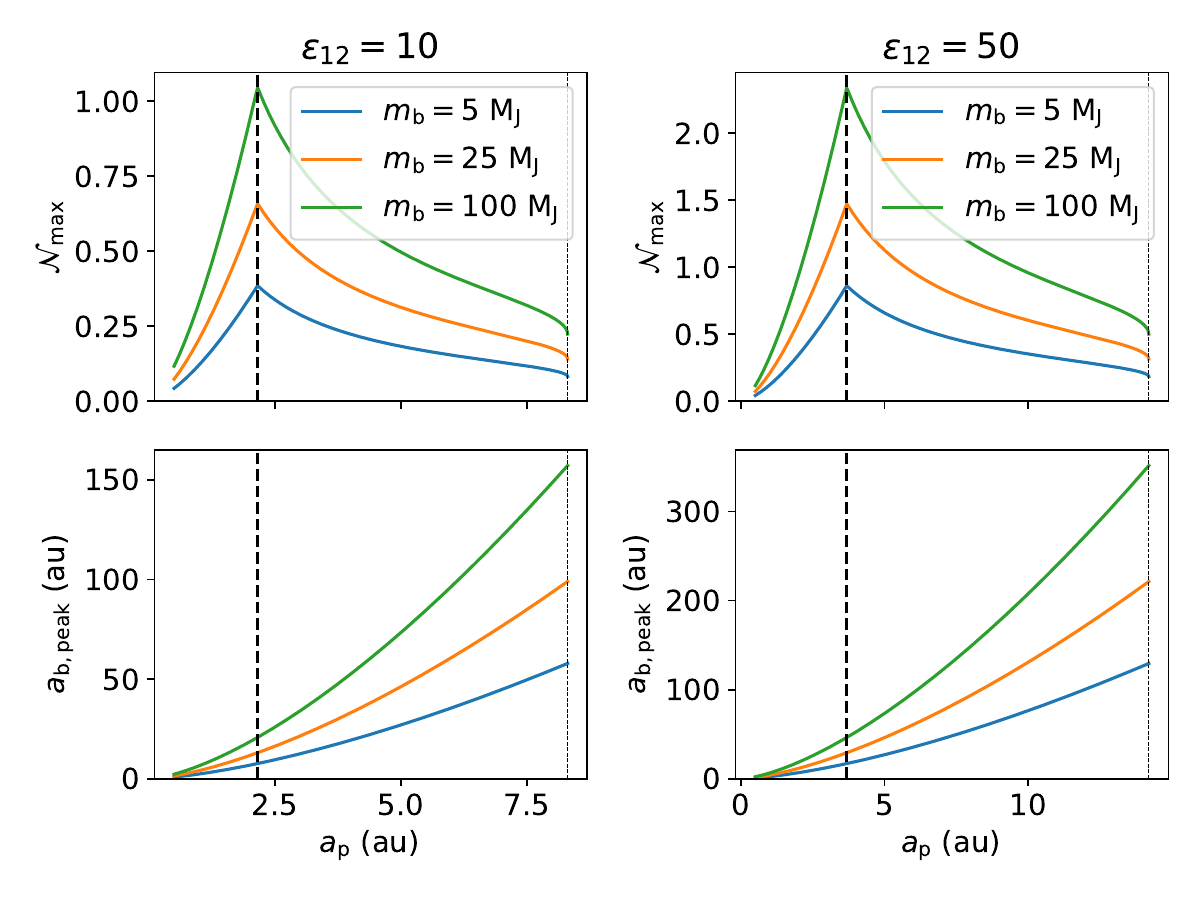}
	\caption{Maximum number of flybys $\mathcal{N}_{\rm max}$ that gives $\Delta I \gtrsim \Delta I_\mathrm{crit} = 0.5^\circ$ (top) and optimal $a_{\rm b, peak}$ (bottom) as a function of $a_{\rm p}$ (equation~\ref{eq:nmax_twop}), for different values of $m_{\rm b}$. Note that $a_{\rm b, peak}$ is chosen so as to maximize the number of flybys, i.e. such that $\varepsilon_{\rm \bar{12}p,b} = 1$ (equation~\ref{eq:abnmax}). The optimal $a_{\rm p}$ is indicated with a dashed vertical line: it corresponds to $\varepsilon_{\rm 12, p} = 1$ and does not depend on $m_{\rm b}$. The maximum $a_{\rm p}$ at around 8 au (left) or 15 au (right) is similarly independent of $m_{\rm b}$ (equation~\ref{eq:amax}). The SEs system is fixed, with $\varepsilon_{12} = 10$ (left) and $\varepsilon_{12} = 50$ (right), $a_{\bar{12}} = 0.5$ au, and the mass of the first companion is fixed to $m_{\rm p} = 1$ $\mathrm{M_J}$. The cluster parameters are $t_\mathrm{cluster} = 20$ Myr, $\sigma_\star = 1$ km/s and $n_\star = 10^3$ pc$^{-3}$. Note that $\mathcal{N}_{\rm max}$ scales with $a_{\bar{12}}^{-1/2}$ and $m_{\rm p}^{-1/3}$, and is proportional to $t_\mathrm{cluster}$, $\sigma_\star^{-1}$, and $n_\star$.}\label{fig:Nflyby_max_vs_ap}
\end{figure*}

\begin{figure*}
	\centering
	\includegraphics[width=0.5\linewidth]{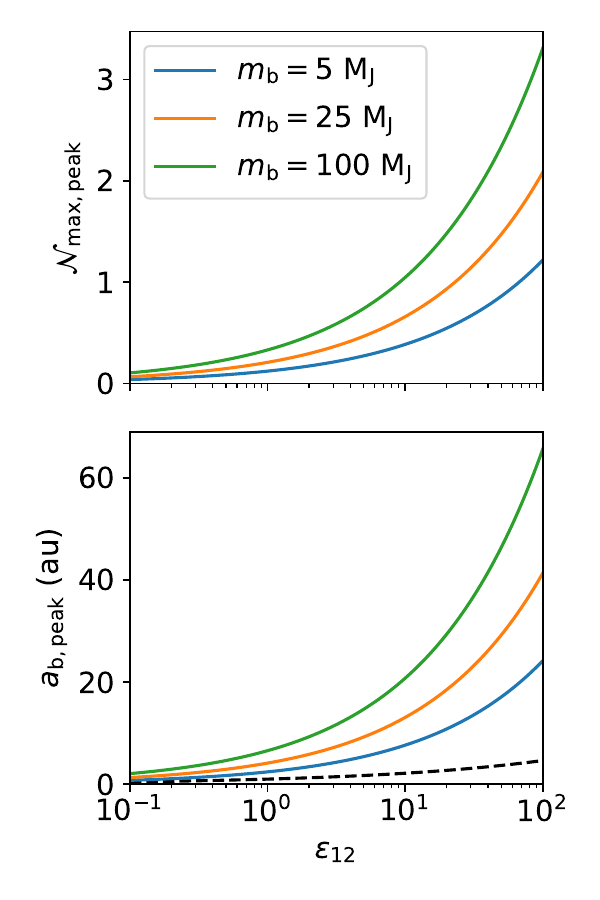}
	\caption{Maximal "peak" number of flybys $\mathcal{N}_{\rm max,peak}$ that gives $\Delta I \gtrsim \Delta I_\mathrm{crit} = 0.5^\circ$ (top) and corresponding $a_{\rm b, peak}$ (bottom) as a function of $\varepsilon_{12}$ (equations~\ref{eq:nmaxopt_twop}--\ref{eq:nmaxopt_twop_bis}), for different values of $m_{\rm b}$. The optimal $a_{\rm p}$ and $a_{\rm b}$ values, $a_{\rm p, peak}$ and $a_{\rm b, peak}$, are set by $\varepsilon_{\rm 12p} = 1$ and $\varepsilon_{\rm \bar{12}p,b} = 1$ (equations~\ref{eq:apnmaxopt}--\ref{eq:abnmaxopt}), so as to maximize the number of flybys. The optimal $a_{\rm p, peak}$ is indicated with a dashed black line on the bottom plot---it does not depend on $m_{\rm b}$. The SEs location is set to $a_{\bar{12}} = 0.5$ au, and the mass of the first companion is fixed to $m_{\rm p} = 1~\mathrm{M_J}$. The cluster properties are $t_\mathrm{cluster} = 20$ Myr, $\sigma_\star = 1$ km/s and $n_\star = 10^3$ pc$^{-3}$, but note that $\mathcal{N}_{\rm max, peak}$ scales with $a_{\bar{12}}^{-1/2}$ and $m_{\rm p}^{1/6}$, and is proportional to $t_\mathrm{cluster}$, $\sigma_\star^{-1}$, and $n_\star$.}\label{fig:Nflyby_max_vs_eps12_twop}
\end{figure*}

Using equation~\eqref{eq:Nflyby} (but applied to the outermost companion $m_{\rm b}$), we find that the number of flybys that can produce $\Delta I > \Delta I_\mathrm{crit}$ is
\begin{align}
	\mathcal{N}(\Delta I > \Delta I_{\rm crit}, a_{\rm b}) &{}=  \mathcal{N}(I_{\rm b} > I_{\rm b, crit}(a_{\rm b}), a_{\rm b}) \nonumber\\ &{}= \mathcal{N}_\mathrm{close}(a_{\rm b}) \times f(I_{\rm b, min} = I_{\rm b, crit}(a_{\rm b})). \label{eq:nflyby_twop}
\end{align}
Figures~\ref{fig:Nflyby_isoinc_twop}--\ref{fig:Nflyby_isoinc_twop_5au} show $\mathcal{N}(\Delta I >\Delta I_{\rm crit}, a_{\rm b})$ as a function of $a_{\rm b}$, for given $m_{\rm b}, m_{\rm p},a_{\rm p}$, and SE properties $\varepsilon_{12}$ and $a_{\bar{12}}$, for the cases of $\varepsilon_{\rm 12, p}\gtrsim 1$ and $\varepsilon_{\rm 12, p}\lesssim1$, respectively.  For a given $a_{\rm p}$, the maximum number of flybys $\mathcal{N}_{\rm max}(a_{\rm p})$ occurs when $\varepsilon_{\rm \bar{12}p, b} \sim 1$. The corresponding $a_{\rm b, peak}$ and $\mathcal{N}_{\rm max}(a_{\rm p})$ are
\begin{align}
	&a_{\rm b, peak} = a_{\rm p} \left(\frac{a_{\rm p}}{a_{\bar{12}}}\right)^\frac{1}{2} \left(\frac{m_{\rm b}}{m_{\rm p}} \right)^\frac{1}{3}\label{eq:abnmax}\\
	&\mathcal{N}_\mathrm{max} = \mathcal{N}_\mathrm{close}(a_{\rm b, peak}) \times f(I_{\rm p, crit}) \propto a_{\rm b, peak}. \label{eq:nmax_twop}
\end{align}
Figure~\ref{fig:Nflyby_max_vs_ap} displays $\mathcal{N}_{\rm max}(a_{\rm p})$ and the corresponding $a_{\rm b, peak}$ as a function of $a_{\rm p}$, for two different $\varepsilon_{12}$ (recall that $\varepsilon_{12}$ characterizes the architecture of the inner SE system; see Fig.~\ref{fig:eps12}). Note that $\mathcal{N}_{\rm max}(a_{\rm p})$ has a peak value, reached when $a_{\rm p}=a_{\rm p, peak}$, corresponding to the transition between regimes where $\varepsilon_{\rm 12,p}\lesssim1$ and $\varepsilon_{\rm 12,p}\gtrsim1$. In other words, for each SE system, there is an optimal companion architecture $(a_{\rm p, peak},a_{\rm b, peak})$ that maximizes the number of effective flybys $\mathcal{N}_{\rm max} = \mathcal{N}_{\rm max, peak}$, which occurs when both $\varepsilon_{\rm 12, p}$ and $\varepsilon_{\rm \bar{12}p, b}$ are approximately equal to $1$. Requiring both coupling ratios to be 1, $a_{\rm p, peak}$ is given by equation~\eqref{eq:apnmax},
\begin{equation}
	a_{\rm p, peak} = (\varepsilon_{12} \tilde{m}_{\rm p})^{1/3}~\mathrm{au},\label{eq:apnmaxopt}
\end{equation}
and $a_{\rm b, peak}$ is given by equation~\eqref{eq:abnmax} with $a_{\rm p} = a_{\rm p, peak}$:
\begin{equation}
	a_{\rm b, peak}= \left(\frac{\varepsilon_{12}}{\tilde{a}_{\bar{12}}}\right)^\frac{1}{2} \left(\tilde{m}_{\rm p} \tilde{m}_{\rm b}^2\right)^\frac{1}{6} ~\mathrm{au},\label{eq:abnmaxopt}
\end{equation}
where $\tilde{a}_{\bar{12}} \equiv a_{\bar{12}}/(1~\mathrm{au})$ and $\tilde{m}_{\rm b} \equiv m_{\rm b}/(1~\mathrm{M_J})$. For a typical system with Earth-mass planets, a cold Jupiter and an outer brown dwarf, we have
\begin{equation}
	a_{\rm b, peak}  \simeq 16~\mathrm{au} \left(\frac{\varepsilon_{12}}{10}\right)^\frac{1}{2} \left(\frac{m_{\rm p}}{1~\mathrm{M_{J}}}\right)^\frac{1}{6} \left(\frac{m_{\rm b}}{50~\mathrm{M_{J}}}\right)^\frac{1}{3} \left(\frac{a_{\bar{12}}}{0.5~\mathrm{au}}\right)^{-\frac{1}{2}} .
\end{equation}
The corresponding $\mathcal{N}_{\rm max, peak}$ is
\begin{align}
	\mathcal{N}_{\rm max, peak} ={} & \mathcal{N}_\mathrm{close}(a_{\rm b, peak}) \times f(\Delta I_\mathrm{crit})\nonumber\\
	\simeq{} &0.8 \left(\frac{\Delta I_{\rm crit}}{0.5^\circ}\right)^{-0.79}  \left(\frac{{\varepsilon}_{12}}{10}\right)^\frac{1}{2} \left(\frac{m_{\rm p}}{1~\mathrm{M_{J}}}\right)^\frac{1}{6} \nonumber\\ &\times \left(\frac{m_{\rm b}}{50~\mathrm{M_{J}}}\right)^\frac{1}{3}   
	\left(\frac{a_{\bar{12}}}{0.5~\mathrm{au}}\right)^{-\frac{1}{2}}  \left(\frac{t_\mathrm{cluster}}{20~\mathrm{Myr}}\right) \nonumber \\
	&\times \left(\frac{M_\mathrm{tot}}{2~\mathrm{M_{\sun}}}\right) \left(\frac{n_\star}{10^3~\mathrm{pc}^{-3}}\right) \left(\frac{\sigma_\star}{1~\mathrm{km/s}}\right)^{-1}\label{eq:nmaxopt_twop} .
\end{align}
This maximal "peak" number of flybys is shown in Figure~\ref{fig:Nflyby_max_vs_eps12_twop}, along with the optimal semimajor axes of the two companions. The result depends on the SE systems through now two parameters, $\varepsilon_{12}$ and $a_{\bar{12}}$. Using equation~\eqref{eq:eps12hat}, we can rewrite equation~\eqref{eq:nmax_twop} to understand the dependence on $\mathcal{N}_{\rm max, peak}$ on various parameters:
\begin{align}
	\mathcal{N}_{\rm max, peak} \simeq {}& 0.8 \left(\frac{\Delta I_{\rm crit}}{0.5^\circ}\right)^{-0.79}  \left(\frac{\hat{\varepsilon}_{12}}{0.3}\right)^\frac{1}{2} \left(\frac{m_2}{10~\mathrm{M_{\earth}}}\right)^{-\frac{1}{2}}\nonumber\\ &\times \left(\frac{a_2}{1~\mathrm{au}}\right)^\frac{3}{2} \left(\frac{a_{\bar{12}}}{0.5~\mathrm{au}}\right)^{-\frac{1}{2}}  \left(\frac{m_{\rm p}}{1~\mathrm{M_{J}}}\right)^\frac{1}{6} \left(\frac{m_{\rm b}}{50~\mathrm{M_{J}}}\right)^\frac{1}{3}   \nonumber \\
	&\times  
	\left(\frac{t_\mathrm{cluster}}{20~\mathrm{Myr}}\right) \left(\frac{M_\mathrm{tot}}{2~\mathrm{M_{\sun}}}\right) \left(\frac{n_\star}{10^3~\mathrm{pc}^{-3}}\right) \left(\frac{\sigma_\star}{1~\mathrm{km/s}}\right)^{-1}.\label{eq:nmaxopt_twop_bis}
\end{align}
Thus, the maximal "peak" number of effective flybys scales linearly with the SEs semi-major axes. Comparing Figures~\ref{fig:Nflyby_max_vs_eps12} and \ref{fig:Nflyby_max_vs_eps12_twop}, we see that adding a second companion strongly increases the number of flybys able to break the co-transiting geometry of the inner SEs. Of course, this increase may be tempered by the lower probability of having a suitable two-companion system instead of an one-companion system. 

\section{Dependence on the mass of the stellar perturber}
\label{sec:mass}

Our calculations in the previous sections assume that the stellar perturber has the same mass as the host star of our planetary system ($M_1 = M_2$). In reality, most of the stars in the Galaxy are less massive than $1~M_{\odot}$, and the impact of a flyby depends on the mass of the perturber $M_2$. On the other hand, for a fixed stellar mass density, if the mass is mostly distributed between lighter stars, we expect $n_\star$ to be higher.

\begin{figure*}
	\centering
	\includegraphics[width=\linewidth]{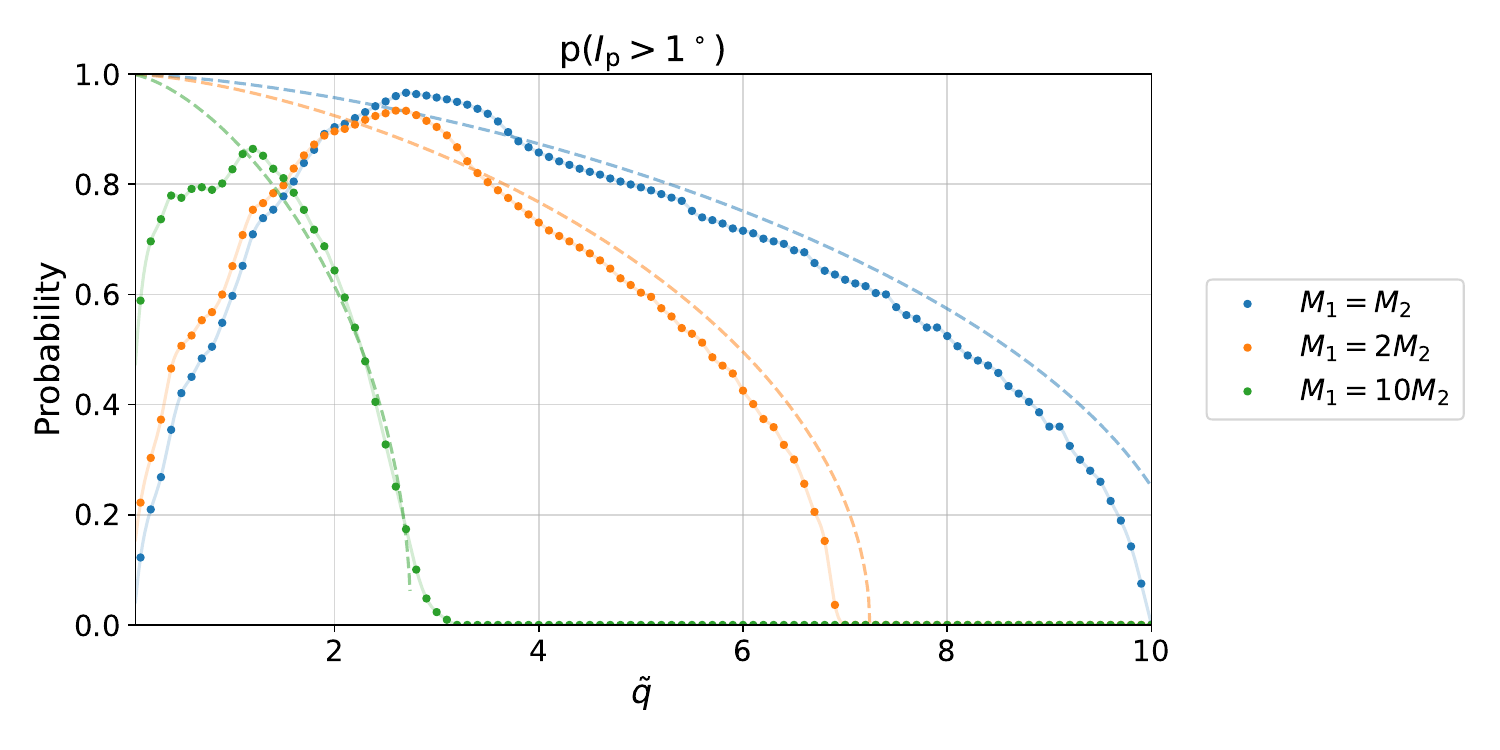}
	\caption{Similar than Fig.~\ref{fig:proba_q_inc}, for $I_{\rm p, min} = 1^\circ$ and different $M_2$ (equation~\ref{eq:rescaling}). For the same distance at closest approach, the effect of the flyby decreases with decreasing perturber mass $M_2$. For small $\tilde{q}$, a decreased mass leads to less configurations becoming unbound, which ultimately increases the probability to have a bound system with small inclination.}\label{fig:proba_q_inc_differentmass}
\end{figure*}

\begin{figure}
 	\centering
 	\includegraphics[width=\linewidth]{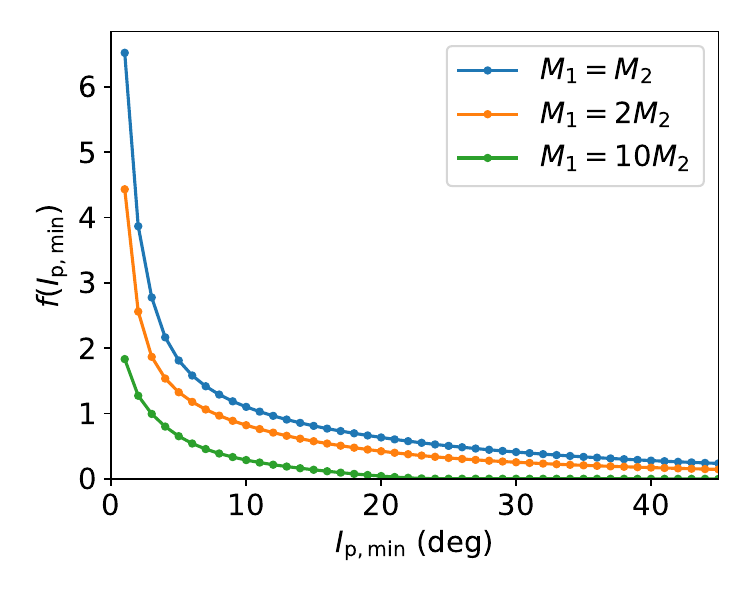}
 	\caption{Same as Fig.~\ref{fig:proba_inc}, for different $M_2$ (equation~\ref{eq:rescaling}).} \label{fig:proba_inc_differentmass}
 \end{figure}

Instead of rerunning all previous simulations with different $M_2$, we can use the $M_1 = M_2$ result to estimate the effect of a perturber with different mass. Inspired from the secular result (see Appendix), all other quantities being the same, the orbital elements of the planet after a flyby with mass ratio $M_2/M_1$ can be derived from the $M_2/M_1 = 1$ results by
\begin{align}
	I_{\rm p} (M_2/M_1)  = \frac{M_2/M_1}{\sqrt{1+M_2/M_1}} I_{\rm p} (M_2/M_1 = 1)\label{eq:rescaling}\\
	e_{\rm p} (M_2/M_1)  = \frac{M_2/M_1}{\sqrt{1+M_2/M_1}} e_{\rm p} (M_2/M_1 = 1).
\end{align} 
This scaling arises because the effect of the  perturbation is proportional to the mass of the perturber, divided by the timescale of the encounter, the latter being proportional to the square root of the total mass $M_{\rm tot}$. We use these estimates to plot Figs.~\ref{fig:proba_q_inc_differentmass} and \ref{fig:proba_inc_differentmass}: The integrated probability $f(I_{\rm p, min})$ is decreased for smaller $M_2$, roughly by a factor of $\sqrt{M_2/M_1}$. Thus, for a fixed stellar number density $n_\star$ in the cluster, the effect of flybys will be decreased by a factor of order a few when we allow for a realistic range of stellar masses, compared to our results presented in previous sections.  On the other hand, if the stellar mass is entirely distributed within $M_2 = M_1/k$ stars, then $n_\star$ is $k$ times higher for the same total mass density. In that case the maximum number of effective encounters $\mathcal{N}_{\rm max} \propto f(I_{\rm p, min}) n_\star$ is actually proportional to $\sqrt{M_1/M_2}$ (e.g., equations~\ref{eq:nmax},\ref{eq:nmax_twop},\ref{eq:nmaxopt_twop}) and will increase for small-mass stars. Thus, we expect that for a given mass density in the cluster, when we consider clusters with a spectrum of stellar masses, the effect of flybys will be increased by a factor of order a few compared to our results presented in previous sections. 

\section{Summary and Discussion}
\label{sec:conclusion}

\subsection{Summary}

In this paper, we have studied a new mechanism for dynamical excitations of Super-Earths (SE) systems following a close stellar encounter. We are motivated by the recently claimed correlation between the multiplicity observed in Kepler transiting systems and stellar overdensities \citep{longmoreImpactStellarClustering2021}. The proposed mechanism consists of stellar flybys exciting the inclinations of one or two exterior companion giant planets (or brown dwarfs), which would then induce misalignment in an initially coplanar SE system. Even a modest misalignment ($\Delta I \gtrsim 0.5^\circ$) could break the co-transiting geometry of the SEs, resulting in an apparent excess of single-transiting systems. We study two cases: in one, the SE system has a cold Jupiter companion at a few au; in the other, the SE system has two companions, a cold Jupiter and a wider planetary or substellar companion. Our results can be rescaled easily with the SE system or companion properties.

To evaluate the probability for a system to experience such an ``effective'' flyby (that can generate $\Delta I \gtrsim 0.5^\circ$ for the SEs), we combine analytical calculation (on stellar encounters and the secular coupling between the SEs and companions) with $N$-body simulations (to evaluate the effects of representative close encounters). Given the large parameter space and related uncertainties, we have attempted to present our results in an analytical or semi-analytical way, so that they can be rescaled easily for different SE and companion properties and stellar cluster parameters.

When the SE system only has one planetary companion, we show that the mechanism is relatively ineffective. If the outer planet is close to the SEs, then stellar flybys will have a small chance of changing its orbital inclination. On the other hand, if the outer planet is farther away, then it will have little impact on the SE dynamics, and on their transiting geometry in particular. This trade-off is shown in Figures~\ref{fig:Nflyby_isoinc} and \ref{fig:Nflyby_isoinc_variations}. For given density, velocity dispersion, and lifetime of the stellar cluster, we can evaluate the expected number of stellar flybys that will disrupt the co-transiting geometry of the SEs as a function of the SE system (characterized by the coupling parameter $\varepsilon_{12}$, equation~\ref{eq:eps12}) and the companion properties (semi-major axis and mass). Fixing all parameters but the companion semi-major axis ($a_{\rm p}$), the number of ``effective'' flybys peaks for an optimal $a_{\rm p}$ which, for typical Kepler SEs, corresponds to a few astronomical units for $m_{\rm p} \sim 1~\mathrm{M_J}$ (see equation~\ref{eq:apnmax}). The corresponding maximum number of effective flybys and its dependencies in all related parameters is displayed in equation~\eqref{eq:nmax}. With only one companion, this flyby-induced misalignment scenario is unlikely to produce statistically significant effect of the SE multiplicity; even when assuming a high cluster density, less than 1 in 10 suitable systems (with an exterior giant planet) will experience the required stellar flyby.

When the SE system has two companions, the effective cross section for stellar flybys increases. The misalignment induced on the outermost (second) companion $m_{\rm b}$ by the flyby can be transmitted to the inner (first) companion $m_{\rm p}$, which would them be able to break the co-transiting geometry of the SEs. A relatively close (small $a_{\rm b}$) second companion would have a strong influence on the inner system (SEs and $m_{\rm p}$) (characterized by the parameter $\varepsilon_{\rm \bar{12}p, b}$, equation~\ref{eq:eps12pb}), and transmit its misalignment $I_{\rm b}$ fully to the first companion (i.e. $I_{\rm b} \sim I_{\rm p}$ for $\varepsilon_{\rm \bar{12}p, b} \gtrsim 1$). On the other hand, a far-out second companion has a better chance of experiencing close stellar flybys, but could have a smaller effect on the inner system. The trade-off is shown in Figures~\ref{fig:Nflyby_isoinc_twop} and \ref{fig:Nflyby_isoinc_twop_5au} for two different locations of the first companion, close to the SEs or farther away. Fixing the properties of the SEs, the number of effective flybys peaks for an optimal combination of $a_{\rm p}$ and $a_{\rm b}$ which, for typical Kepler ``SE+cold Jupiter systems'', corresponds to a few AU and tens of AU, respectively, for substellar second companions. The corresponding number of flybys and its dependencies on various parameters is displayed in equation~\eqref{eq:nmaxopt_twop}. The optimal architecture (semi-major axes of the two companions as a function of their masses and the SE parameters) is described in equations~\eqref{eq:apnmaxopt}--\eqref{eq:abnmaxopt} (see Figures~\ref{fig:Nflyby_max_vs_ap}--\ref{fig:Nflyby_max_vs_eps12_twop}). Our calculations show that the number of effective flybys (that can generate $\Delta I > 0.5^\circ$ for the SEs) increases significantly when a second companion is added, suggesting that most initially coplanar SE systems with two companions in stellar overdensities will experience a flyby that will misalign their orbits, and thus break the co-transiting geometry. 

\subsection{Discussion}

In this paper, we have adopted some simplifications in deriving the analytical estimates of the effect of flybys on SE systems with exterior companions. One of the simplifications that we made was to ignore the encounters with binary stars. Tight binaries with separation $a_{\rm binary} \ll a_{\rm p}$ are expected to behave similarly to a single stellar perturber. On the other hand, only one component will play a role in encounters with large-separation binaries ($a_{\rm binary} \gg a_{\rm p}$). For intermediate cases, more exploration is needed \citep[see e.g.][]{liEncountersInvolvingPlanetary2020}.

Our calculations show that SE systems with only one outer companion are unlikely to be much impacted by the stellar flybys, while systems with two companions are likely to be impacted (depending on various parameters). However, SE systems with two companions may be rarer than their one-companion counterparts. Ultimately, further information on the correlation between inner SE systems and exterior planets/substellar companions will be needed to determine the relative prevalence of the mechanisms studied in this paper. Moreover, a deeper knowledge on the properties of stellar birth clusters (lifetime and density in particular) and their relation to the observed overdensities characterized in \cite{winter_stellar_2020} and \cite{longmoreImpactStellarClustering2021} is needed to draw firm conclusions on the role of the flyby-induced misalignment scenario in multi-planet systems.

\section*{Acknowledgements}

We thank the referee, A. Mustill, for useful comments that have improved this paper. This work has been supported in part by the NSF grant AST-17152 and NASA grant 80NSSC19K0444. We made use of the \textsc{python} libraries \textsc{NumPy} \citep{harrisArrayProgrammingNumPy2020}, \textsc{SciPy} \citep{virtanenSciPyFundamentalAlgorithms2020}, and \textsc{PyQt-Fit}, and the figures were made with \textsc{Matplotlib} \citep{hunterMatplotlib2DGraphics2007}.

\section*{Data Availability}

The output of the $N$-body flyby simulations (Section~\ref{sec:flyby}) can be found on \url{https://github.com/LaRodet/FlybySimulations.git}. All the figures can be directly reproduced from this dataset and from the equations.



\bibliographystyle{mnras}
\bibliography{Biblio} 




\appendix

\onecolumn

\section{Secular perturbation on the inclination by a parabolic flyby}

\begin{figure}
	\centering
	\includegraphics[width=0.5\linewidth]{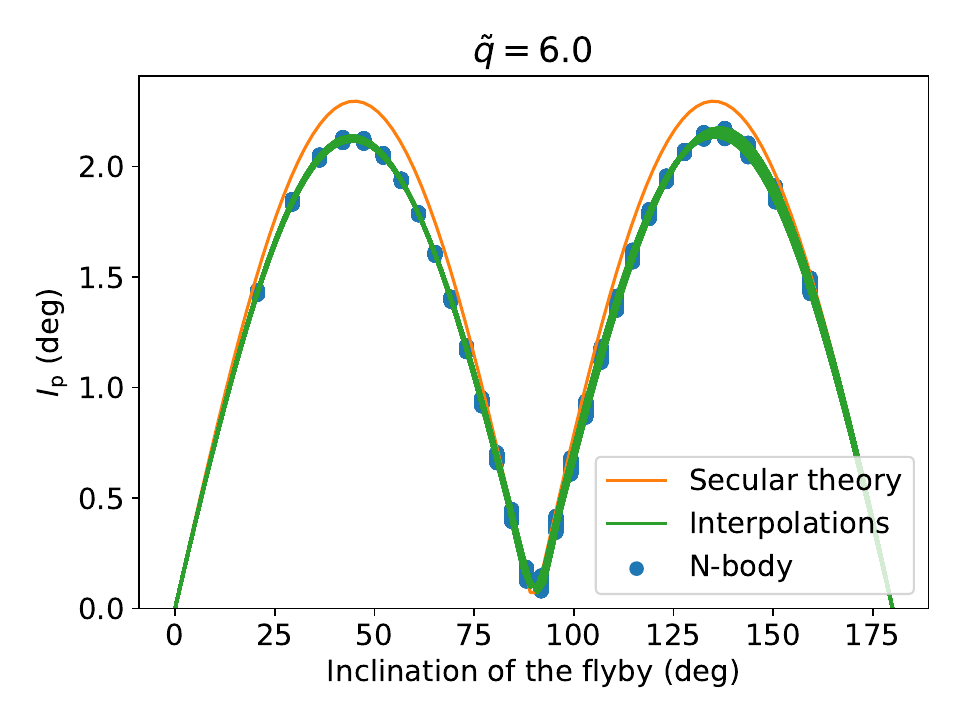}
	\caption{Inclination of the planet after the flyby, from the secular theory (orange, equation~\ref{eq:probasecular}) and $N$-body simulations (clustered blue dots). Each of the 30 simulated inclinations groups $20\times 20$ simulations, with various $\omega$ and $\lambda_{\rm p}$. The green lines are the interpolation of $I_{\rm p}$ over the inclination of the flyby for each couple ($\omega$,$\lambda_{\rm p}$). These results and interpolations are used to produce Figure~\ref{fig:proba_q_inc}.}\label{fig:ip}
\end{figure}

The probability $p\left(I_{\rm p} > I_{\rm p, min}\right)$ for a planet $m_{\rm p}$ to gain an inclination $I_{\rm p} > I_{\rm p, min}$ after a parabolic flyby can be computed in the secular framework, for $\tilde{q} \equiv q/a_{\rm p} \gtrsim 3$. Let us first compute the evolution of the unit angular momentum vector $\vec{\hat{L}_{\rm p}}$ of an initially circular planet orbit (with semi-major axis $a_{\rm p}$ and mean-motion $n_{\rm p}$). Using the secular approach, averaging over the planet's orbit, we get (Eq.~28 in \cite{liuBlackHoleNeutron2018}):
\begin{equation}
	\dv{\vec{\hat{L}_{\rm p}}}{t} = -\frac{3 n_{\rm p} M_2 a_{\rm p}^3}{2M_1 R^3} \left(\vec{\hat{L}_{\rm p}}\cdot\hat{\vec{R}}\right)\left(\vec{\hat{L}_{\rm p}}\times\hat{\vec{R}}\right),
\end{equation}
where $\vec{R}$ represents the instantaneous position of $M_2$ relative to $M_1$, and $\hat{\vec{R}} = \vec{R}/R$. We choose the $xy$ plane as the plane of the parabolic stellar encounter, with the $x$ axis pointing towards the periastron. We assume that $\vec{\hat{L}_{\rm p}}$ remains close to its initial value throughout the flyby. Integrating $\dv{\hat{L}_{\rm p}}{t}$ along the parabolic trajectory of the flyby, we find that the change in $\vec{\hat{L}_{\rm p}}$ is given by
\begin{equation}
	\vec{\delta \hat{L}_{\rm p}} = \frac{3 \pi M_2}{8 \sqrt{2 M_1(M_1+M_2)}} \left(\frac{a_{\rm p}}{q}\right)^\frac{3}{2} \sin(2 i) \left(\begin{array}{c} \cos\Omega
		\\ \sin\Omega \\0
	\end{array}\right) ,
\end{equation}
where $i$ and $\Omega$ are respectively the inclination and longitude of the node of the initial orbit of the planet. Thus, the change in the planet's inclination is
\begin{align}
	I_{\rm p} &{}\simeq |\vec{\delta \hat{L}_{\rm p}}| =  \frac{3\pi}{8} \frac{M_2}{\sqrt{2 M_1 (M_1 + M_2)}} \left(\frac{a_{\rm p}}{q}\right)^{\frac{3}{2}} \sin(2i)\nonumber\\
	&{} \equiv I_{\max}(\tilde{q}) \sin(2i).
\end{align}
This expression is compared to the N-body results in Fig.~\ref{fig:ip} for $\tilde{q} = 6$. The dependence on $i$ is correctly predicted, but the amplitude is slightly overestimated (by about $10\%$). This is due in part to the N-body setup, where the eccentricity of the flyby is $1.1$ rather than 1. This slight overestimate of the amplitude can lead to a larger discrepancy of the probability to generate $I_{\rm p} > I_{\rm p, min}$, if this inclination threshold is close to $I_{\rm max}(\tilde{q})$. In fact, we can easily compute the theoretical probability assuming the flyby has a uniform distribution in $\cos i$:
\begin{align}
	p\left(I_{\rm p} > I_{\rm p, min}\right) ={}& \frac{1}{2}\int_{I_{\rm p} > I_{\rm p, min}} \sin(i) di\nonumber\\
	={}& \int_{i > \frac{1}{2}\arcsin(\frac{I_{\rm p, min}}{I_{\rm max}(\tilde{q})})}^{i < \frac{\pi}{2} - \frac{1}{2}\arcsin(\frac{I_{\rm p, min}}{I_{\rm max}(\tilde{q})})} \sin(i) di\nonumber\\
	={}& \cos\left[\frac{1}{2}\arcsin(\frac{I_{\rm p, min}}{I_{\rm max}(\tilde{q})})\right] - \sin\left[\frac{1}{2}\arcsin(\frac{I_{\rm p, min}}{I_{\rm max}(\tilde{q})})\right].\label{eq:probasecular}
\end{align}
This probability is indicated by the dashed lines in Fig.~\ref{fig:proba_q_inc}.

\bsp	
\label{lastpage}
\end{document}